# Urban Scaling in Denmark, Germany, and the Netherlands
*Relation with Governance Structures*

Anthony F.J. van Raan

Centre for Science and Technology Studies
Leiden University
Kolffpad 1
P.O. Box 905
2300 AX Leiden, The Netherlands
vanraan@cwts.leidenuniv.nl

*Abstract*

*We investigate the socioeconomic urban scaling behavior in three European Union countries: Denmark, Germany, and the Netherlands. In the case of Denmark our analysis relates to all 98 municipalities. We examine the scaling of larger cities, municipalities within the Copenhagen agglomeration, and municipalities in rural areas. We also distinguish between municipalities with high and low centrality. Superlinear scaling of the gross urban product with population size is found in all cases, with exponents between 1.14 and 1.24. In Germany we distinguish between major cities of which the surrounding urban region belongs to the municipality of the city, the 'kreisfreie Städte' (in total 106), and 'Kreise', i.e., regions around smaller cities consisting of several municipalities (in total 296). A striking finding is that the scaling exponents differ substantially from one region to another. We find in most cases a significant superlinear scaling with exponents up to 1.33. Our analysis shows that urban regions with one municipality perform significantly better than urban regions with fragmented governance structures (more than one municipality). A strong relation is found between the measured residuals of the scaling equations and the socio-economic position of a cities assessed with a set of different indicators. We also investigate the relation between scaling of the Kreise and measures of centrality, including the Zipf-distribution. For Germany as well as for Denmark we find that urban scaling is related to generative and not to distributive processes. For the Netherlands we focus on the group of the major cities (in total 21) with their agglomerations and on all 60 municipalities in the Province of Zuid-Holland. We examined the scaling of municipalities within larger urban areas in the Netherlands. In all cases significant superlinear scaling is measured with exponents up to 1.28. Our earlier observation that one-governance urban areas perform better than multi-governance urban areas is confirmed and this is in line with the above mentioned findings for Germany. This leads to challenging conclusions about the importance of a one-municipality instead of a multi-municipality governance in major urban regions. A coherent governance of major cities and their agglomerations may create more effective social interactions which reinforce economic and cultural activities generating a substantial wealth benefit.*

# 1. Introduction and Political Context

In recent years there is a rapidly growing interest in the role of cities in our national and global society. Cities are regarded as the main locations of social, economic, cultural and innovative activity [1]. Recent research on urban metrics shows a *more than proportional* (superlinear) increase of the socio-economic performance of cities measured by the gross urban product (GUP) with increasing population [2, 3, 4]. In this *urban scaling* the dependence of GUP (denoted as $G(P)$) on population ($P$) is given by the *power law* relation:

$$G(P) = aP^\beta \tag{1}$$

where the coefficient $a$ and the urban scaling exponent $\beta$ follow from the empirical data. In most cases the value of $\beta$ is between 1.10 and 1.20. This implies that a city twice as large (in population) as another city can be expected to have a factor of about 2.15 larger socio-economic performance (in terms of the gross urban product). This urban scaling behavior is not restricted to socio-economic variables such as GUP: it is found for human interactions in general and for knowledge production activities [5, 6, 7, 8] in cities. Indicators representing these activities appear to scale nonlinearly with the number of inhabitants of cities and urban agglomerations. Similar scaling is found for other complex systems such as universities [9]. The basis of this scaling behavior is provided by the theory of complex, adaptive systems [10]. Networked structures reinforce nonlinearly as the system grows, particularly more than proportional, i.e. superlinearly, described by a power law [11].

A simple way to understand this phenomenon is by realizing that the number of nodes increases *linearly* whereas the number of links between the nodes increases *superlinearly* with the growth of a network. The nodes in the urban complex system are the inhabitants, social and cultural institutions, centers of education and research, firms, etcetera. The links between these (clustered) nodes are crucial for new developments, reinforcement of urban facilities, and innovation. Because they increase superlinearly, the socio-economic strength of the city increases more than proportional with increasing size of the city. For an extensive discussion of the theoretical basis of urban scaling we refer to [11]. Evidently, the relation between urban scaling (which is a phenomenon at the meso-level) and dynamic processes in urban systems, for instance the concentration of business companies and professions, mobility, and other forms of traffic relations (which are processes at the micro-level) are important and could provide further understanding of scaling.

The US research on urban scaling is about urban areas (MSA's, metropolitan statistical areas) that have grown autonomously to a specific number of inhabitants, regardless of the formal boundaries of municipalities within an urban area. This is particularly the case for a recent study on urban scaling in Europe [12] where cities are defined as 'functional cities' on the basis of recent OECD-EU definitions of large metropolitan areas [13]. Urban scaling analysis is based on data of cities of different population size within the same time window, and thus urban scaling means a synchronic, 'static' measurement that has a predicting value for what happens with socioeconomic variables if, for instance, a city (i.e., urban area) doubles in population in the course of time. This is, of course, different from a situation in which a city defined as a municipality and being the central city of the urban agglomeration, doubles in population by a formal reorganization of all municipalities within the urban area into one new municipality. Nevertheless it is probable that after some time the newly formed city should meet the scaling values as predicted by its new size of population. But crucial is here the interesting policy question: would these scaling values for the doubled population ('created' by municipal reorganization) not already be attained for the urban agglomeration as a whole, simply because the urban agglomeration regardless of the formal municipal boundaries already has this double population?

Our previous study on the urban scaling of cities in the Netherlands [14] focused precisely on this problem by analyzing the scaling behavior of major cities in three structural and administrative modalities: the municipality of the central city, their urban agglomerations and their urban areas. In all three modalities superlinearity with power-law exponents of around 1.15 was found. But remarkably, agglomerations and urban areas underperform if we compare for the same size of population an agglomeration or urban area with a city as one municipality. In other words, an urban system as one formal municipality (one-governance) performs better as compared to an urban agglomeration (multi-governance) with the same population size. Moreover, further analysis suggested that cities with a municipal reorganization recently and in the past decades have a higher probability to perform better than cities without municipal restructuring.

A recent OECD study [15] supports our earlier finding. The authors conclude that in line with previous literature, their analysis confirms that city productivity tends to increase with city size. But even more important, on the governance side the authors find that cities (urban regions) with fragmented governance structures tend to have lower levels of productivity. For instance, for a given population size, a metropolitan area with twice the number of municipalities is associated with around six percent lower productivity.

The urban scaling phenomenon is important for new insights into and policy for urban development and, particularly, for municipal reform. This may concern urban agglomerations where municipal reform means an enlargement of the municipality of the central city by discontinuation of the municipalities of the suburbs, and of smaller municipalities that are merged into one new municipality. Different from the usual focus on measuring the effect of municipal reform on expenses, the urban scaling phenomenon relates to the expected socio-economic profits. Possible effects per medium-sized city could amount to hundreds of millions of euros which means thousands of jobs per year and per urban area. Furthermore, the interpretation of urban scaling laws is important in the discussion on models of urban growth, structure and optimal size of cities and their regions [16, 17].

The main goal of the study presented in this paper is (1) to examine urban scaling in *different national systems* and, with that, to find out to what extent urban scaling depends on characteristics of these national systems; (2) to find further empirical evidence of the above mentioned differences in scaling of *one-governance versus multi-governance urban systems*; and (3) and to relate deviations from the average scaling behavior to *socioeconomic characteristics* of urban areas. The structure of this paper is as follows. We describe our data and method to investigate the urban scaling behavior in three European Union countries, Denmark, Germany, and the Netherlands. The accents of our study differ for each of these three countries. For Denmark, the urban scaling is analyzed against the background of a nation-wide municipal reform. In the case of Germany, we study urban scaling in different parts of the country with considerable differences in economic strength. Furthermore we focus on cities that form with the surrounding urban area one municipality ('kreisfreie Städte'), on administrative units that consist of several municipalities ('Kreise'). Moreover, for Denmark as well as for germany we examine the effects of centrality on urban scaling. In the case of the Netherlands our analysis concerns major cities with agglomerations that consist of a number of municipalities, here urban scaling is investigated for the central cities of the agglomeration, and for the agglomeration as a whole. Moreover, scaling is analyzed for all municipalities in the most populous province, South Holland. Also urban scaling within urban agglomerations is measured. Finally we discuss the results and we conclude the paper with a discussion of the policy implications.

## 2. Denmark

2.1 Data

Denmark with its nearly 6 million inhabitants is administratively divided in five regions and these regions are subdivided into municipalities. In 2007 the number of municipalities was reduced from 271 to 98 [18]. This municipal reform, however, mostly involved the smaller, more rural municipalities. Given the population size of Denmark, the number of major cities is restricted: seven municipalities have more than 100,000 inhabitants and four of these are municipalities of which the central city itself has a population of more than 100,000 (Copenhagen, Aarhus, Odense Aalborg). These major cities -except for Copenhagen- had earlier municipal reforms in the 1970's and 1980's and only Aalborg had a municipal reform in 2007. Remarkably, there has been no municipality reform whatsoever of the Copenhagen agglomeration (formally called the Capital Region of Denmark). Of the 98 Danish municipalities 26 are in the Copenhagen agglomeration (including Copenhagen). We collected for Denmark the following data: (1) gross urban product and (2) population (number of inhabitants) of all municipalities in Denmark for the period 1997-2015[1]. In addition we collected for all Danish municipalities their surface areas[2] and the population of the main town or city of the municipality[3]. The surface areas are used to calculate the population density in a municipality, and number of inhabitants of the main town or city within a municipality is used to calculate the centrality of the municipality.

2.2 Scaling Analysis: Total Set and Subsets of Municipalities

We first analyzed the scaling of the gross urban (municipal) product as a function of population for the entire set of 98 municipalities and show the results in Fig 1. We find that the scaling exponent is 1.16. The regression line represents the expected values of the gross urban (municipal) product and, as can be expected, the real values of the gross urban (municipal) product for the municipalities deviate from the expected values. These differences are called residuals; for the calculation of residuals we refer to the Appendix. We will discuss the meaning of residuals in urban scaling analyses later in this paper.

In some cases the deviation of an individual city or municipality from the regression line can be remarkably large. In the measurement of urban scaling the effect of such 'outliers' is an important issue. Indeed, as the figure suggests, the measured exponent is sensitive to outliers. There are several statistical tests to find out whether a data point can be considered as an outlier, but in most cases typical outliers are immediately visible in the scaling figure. A striking example is Billund (indicated with a circle in the upper part of Fig 1), a small municipality of about 26,000 inhabitants but with a GUP four to five times higher than other municipalities of similar size. The reason for this exceptionally high GUP is obvious: Billund is the home town of Legoland, the largest tourist attraction in Denmark with two million visitors per year. Furthermore, Billund International Airport is the second largest airport in Denmark. Billund has by far the largest relative increase of GUP of all Danish municipalities in the last ten years. If we remove Billund as an exceptional outlier the exponent is 1.18.

---

[1] The data on Gross Domestic Product (in current prices, previous years prices and 2010-prices, chained values) and population at municipality level were provided by Danmarks Statistik (Statistics Denmark) www.dst.dk.  Data are compiled and analyzed in file DK-UrbScal.xlsx.
[2] Retrieved from https://en.wikipedia.org/wiki/List_of_municipalities_of_Denmark.
[3] For instance, see https://en.wikipedia.org/wiki/Esbjerg_Municipality.

Other outliers are two municipalities within the Copenhagen agglomeration, Ballerup and Glostrup. For instance Glostrup has twice as many employees as compared with municipalities of similar size because several major international companies are located in Glostrup. Ballerup has the highest concentration of companies and jobs after Copenhagen. Removing all three outliers results in a scaling exponent of 1.20. These observations clearly show the sensitivity of the measured scaling behavior to the parameter values of individual cities/municipalities.

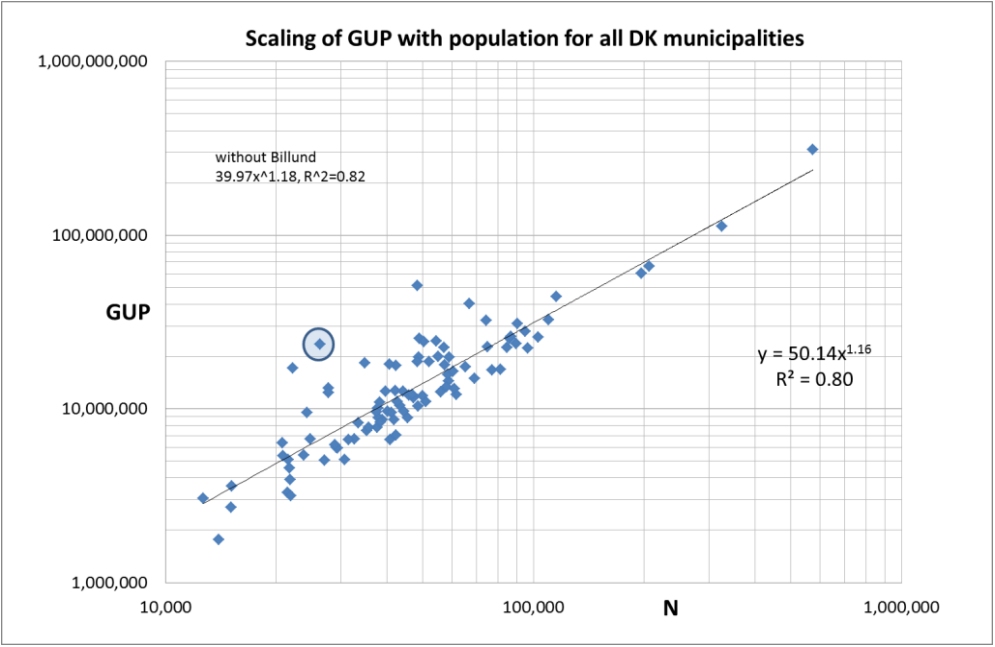

Fig 1. Scaling of GUP (in 1,000 DDK) for all Danish municipalities (data: average 2013-2015).

In the first scaling analysis shown in Fig 1 all municipalities of Denmark were involved. In a further step we investigate the scaling behavior for specific *subsets* within the total set of all municipalities. To the best of our knowledge this has never been done so far. This is important because of the crucial question: do the urban scaling exponents depend on specific characteristics of the set of cities/municipalities for which the scaling is measured? We focus on three characteristics: population, density of a group of cities, and centrality in terms of density within a municipality. Urban scaling of the following subsets is measured.

(1) the municipalities with a population over 50,000; this means a subset of the larger municipalities;

(2) the municipalities within the Copenhagen agglomeration; this means a subset of municipalities that are very close together;

(3) municipalities with centrality <0.50; this means a subset of mostly the smaller, more rural municipalities.

We use a simple but workable measure of centrality: the ratio of the population of the main town or city in the municipality to the total population of the municipality. In municipalities around larger cities at least half of the total population lives in the central city. Thus, for the typical urban areas the centrality will be above 0.50. Such urban areas are therefore mostly 'monocentric'. In rural areas, however, the main towns of municipalities are often small with less than 10,000 inhabitants and the remaining inhabitants are divided over a larger number of smaller towns within the municipality. As a consequence, the centrality of rural area municipalities is smaller than 0.50. In such a situation the municipalities are more 'polycentric'. In Fig 2 we present the results for the first subset, all municipalities (n=37) with a population over 50,000. We find that the scaling exponent is 1.14 [4].

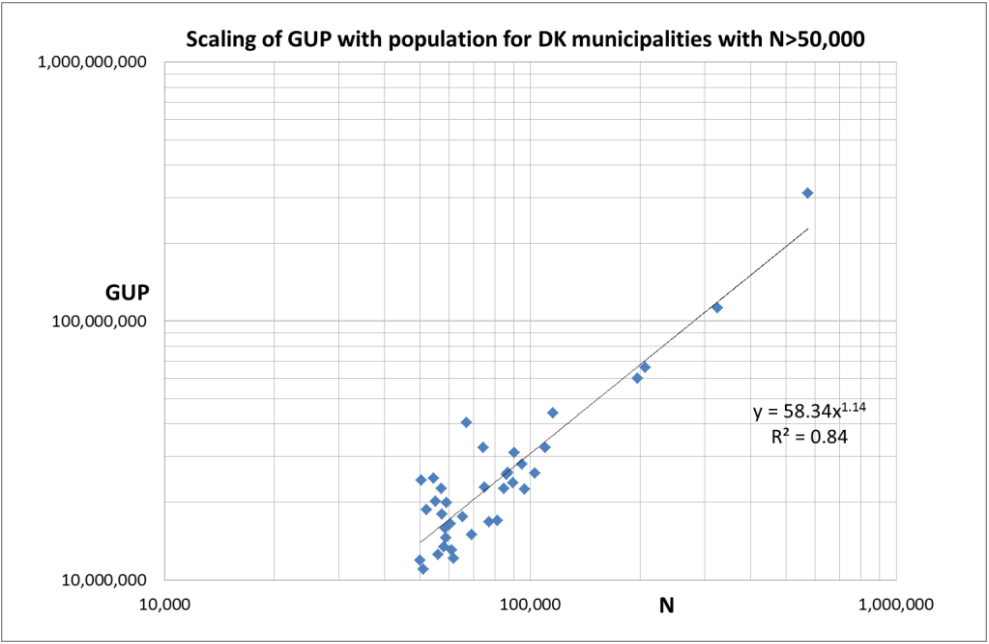

*Fig 2. Scaling of GUP (in 1,000 DDK) for all Danish municipalities above 50,000 inhabitants (data: average 2013-2015).*

Next, we analyze the scaling of the second subset: all municipalities (n=28) within the Copenhagen agglomeration, see Fig 3 for a map. As shown in Fig 4, we find a scaling exponent 1.24. This means that also *within a compact urban area* the autonomous municipalities exhibit scaling behavior. Fig 3 also shows the outlier position of Glostrup and Ballerup. As discussed earlier, these municipalities are part of the compact capital urban region which is the main Danish economic center. Because these agglomeration municipalities are located at the low left side of the regression line they 'lift' as it were the regression line which causes a lower exponent if they are included, and a slightly higher exponent if they are excluded (1.28).

---

[4] We refer to [14] for an extensive error discussion. On the basis of these calculations we estimate that the 95% confidence interval is within +/- 0.05 around the measured scaling exponent.

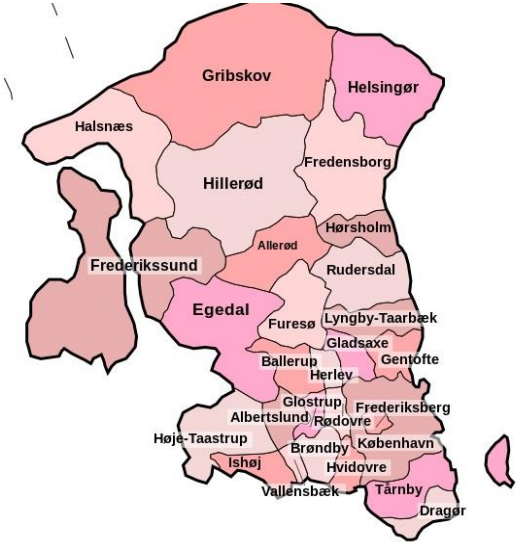

*Fig 3. Map of the Copenhagen agglomeration municipalities[5].*

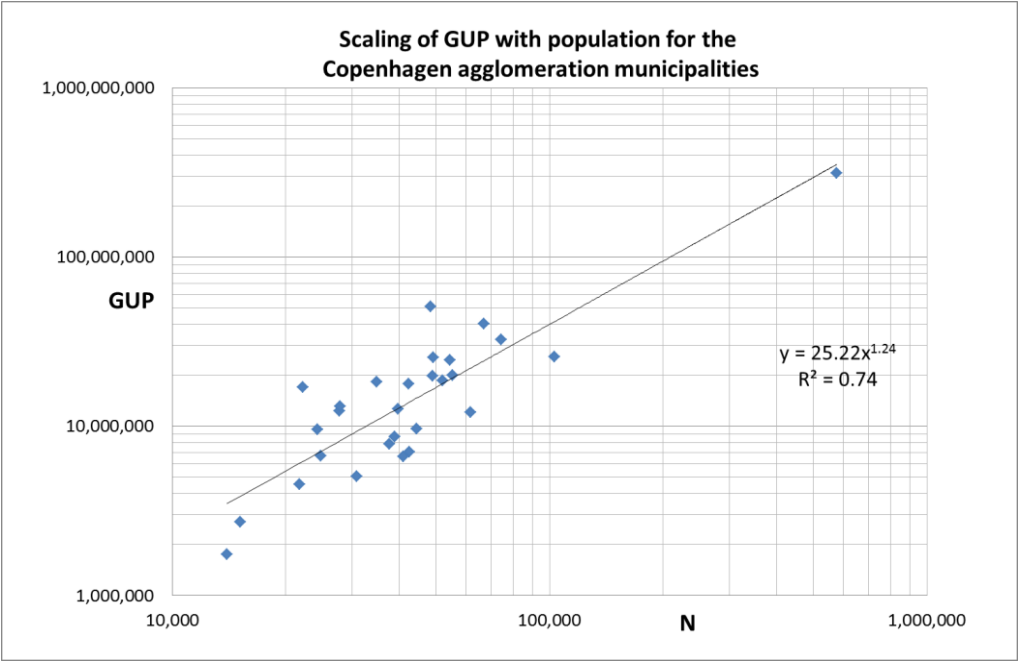

*Fig 4. Scaling of GUP (in 1,000 DDK) for all municipalities within the Copenhagen agglomeration (data: average 2013-2015).*

We see that in the Copenhagen agglomeration the power law exponent is somewhat higher than in the case of all municipalities (Fig 1) and in the case of the municipalities above 50,000 inhabitants (Fig 2). This raises the question whether this higher exponent of urban scaling is caused by the fact that the municipalities in the Copenhagen agglomeration are very close together, and thus the whole subset can be seen as a strongly interconnected network of municipalities. If we leave out Copenhagen municipality, the scaling exponent is even higher, 1.30. An explanation could be that Copenhagen municipality performs relatively less within the entire agglomeration. This raises the question whether this finding has consequences for the political strategy to abandon municipal reform of the Copenhagen agglomeration. Because we have in Denmark only one large urban agglomeration (Copenhagen) we investigate the urban scaling within several major urban agglomerations the Netherlands, see Section 4.2.3.

---

[5] *https://en.wikipedia.org/wiki/Capital_Region_of_Denmark#Municipalities_of_Region_Hovedstaden.*

The third subset relates to centrality. In order to focus on the more rural municipalities we measure the scaling of the Danish municipalities, not within the Copenhagen agglomeration, with centrality *smaller* than 0.50 (n=49, half of all Danish municipalities, we excluded Billund). The result is shown in Fig 5. We find a scaling exponent 1.14, not very different from the scaling exponent of all Danish municipalities (without Billund 1.16).

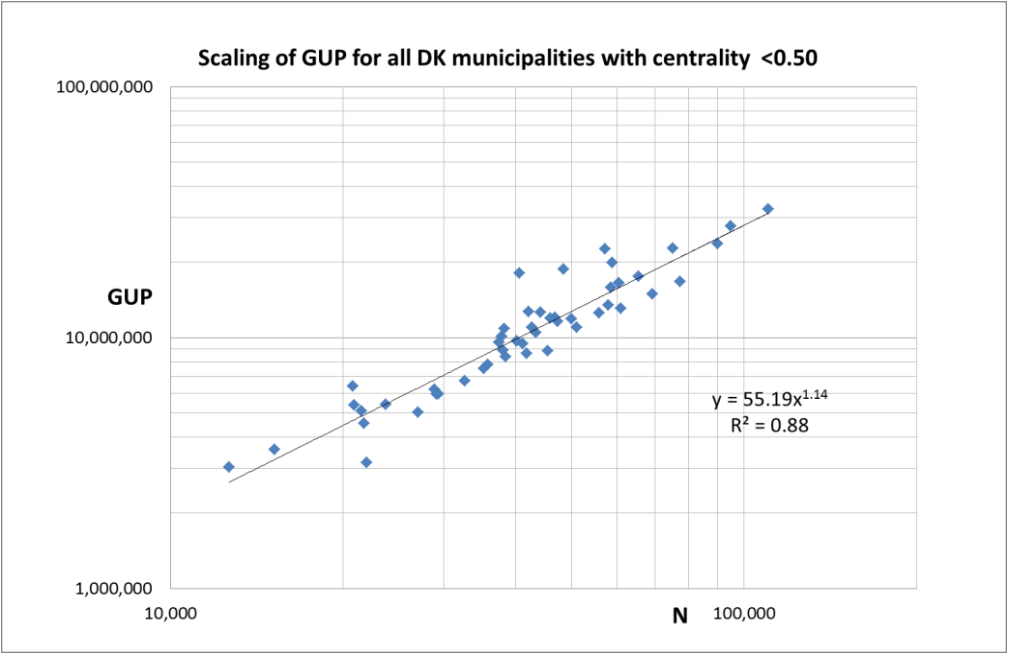

*Fig 5. Scaling GUP (in 1,000 DDK) for all Danish municipalities with centrality <0.50 and not in the Copenhagen agglomeration (data: average 2013-2015).*

Finally, we measured the scaling behavior of all Danish municipalities with centrality *larger* than 0.50 which concerns most of the larger cities. In Fig 6 we show the results without the Copenhagen agglomeration. We find a scaling exponent of 1.23 (n=18). If we include the Copenhagen municipalities (almost all of them have centrality >0.50) the scaling exponent is lower, 1.14. In this case we cover practically all municipalities with more than 50,000 inhabitants and indeed the exponent is the same as shown in Fig 2.

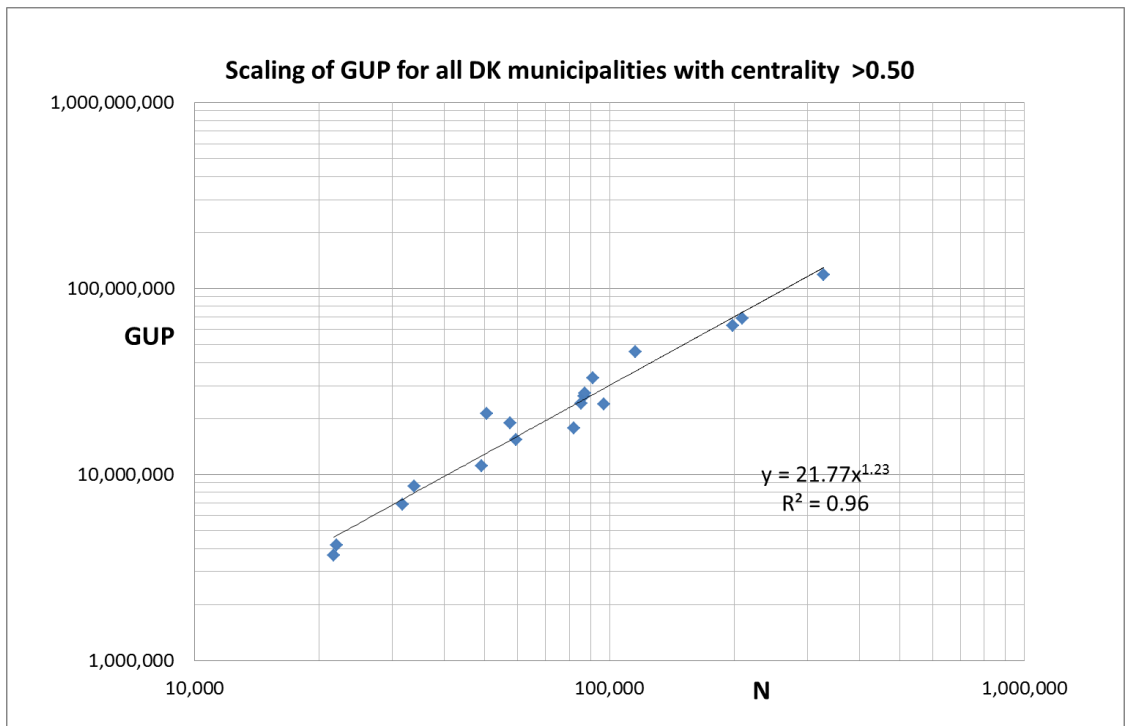

*Fig 6. Scaling of GUP (in 1,000 DDK) for Danish municipalities with centrality >0.50, we exclude the Copenhagen agglomeration (data average 2013-2015).*

Our results show that scaling behavior at the level of municipalities is present in all investigated contexts: major cities, rural areas, and within the large urban agglomeration of the Danish capital region.

## 2.3 Generative versus Distributive Processes

It is often claimed that the non-linear increase of GUP as a function of population size benefits the larger cities at the cost of smaller cities. Thus, scaling would be a result of *distributive* processes instead of *generative* processes. These latter imply that cities of all sizes benefit. It is not that difficult to find out what is going on. If we calculate the increase of GUP over a period of, for instance 10 years, there must be a clear positive dependence on population size if scaling would involve distributive processes. Fig 7 shows that this is not the case. For all Danish municipalities (also for the subset of municipalities with centrality <0.50) we find no significant dependence of the ratio GUP(2014)/GUP(2005) on population size. We conclude that urban scaling is related to generative processes.

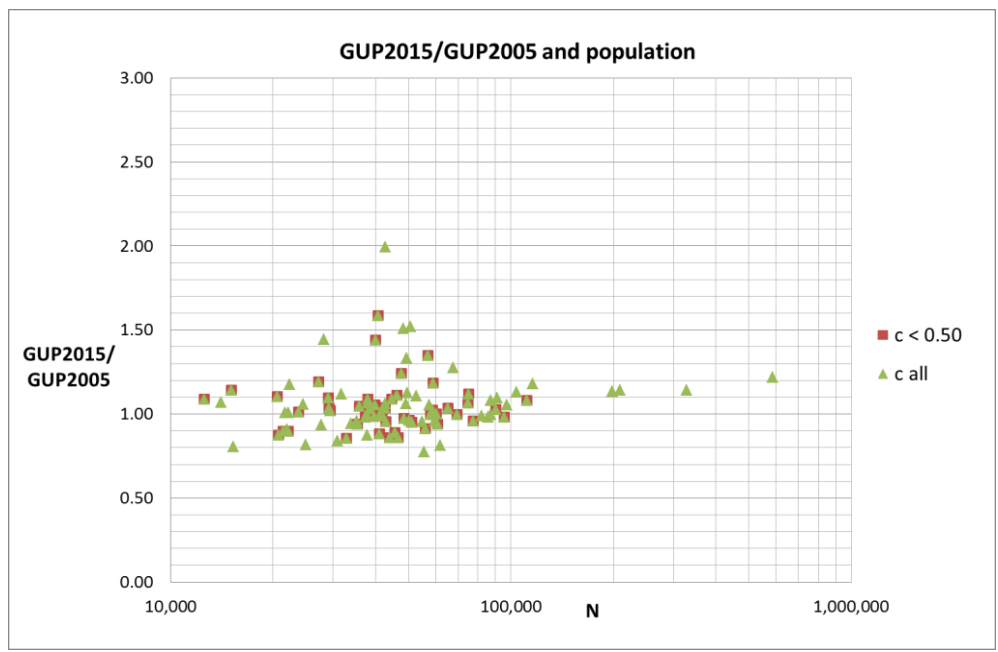

*Fig 7. Increase of GUP for the Danish municipalities between 2005 and 2015 (price index: 2010) as a function of population size.*

## 2.4 Discussion of Research Related to Municipal Reforms

There is a vast amount of literature on municipal reform (also called 'municipal amalgamation' or 'municipal mergers') but to the best of our knowledge there is no work on municipal reform that take urban scaling, i.e., the disproportional increase of the gross urban product as a function of population, into account. Municipal reform literature focuses on the cost-side and not on the profit. Danish work [18, 19] focuses on the financially opportunistic behavior in terms of 'last minute' spending at the end of the budget year by the 'old' municipalities in the years before the Danish municipal reform in 2007 took place. Similar 'free-riding' spending behavior was found for municipal reform processes in Sweden [20, 21]. In this context it was concluded that municipal reform will only be successful if long-term benefits of municipal reform outweigh the short-term costs associated with them [22]. Also for the Finnish municipal reform around 2008 the free-riding behavior was studied [23] and the authors suggest financial constraints during the process of merging municipalities to mitigate this problem. In a further Danish study [24] it is found that an increasing municipal size through municipal reform initially has an adverse effect on fiscal outcomes, but there were improvements after four to five years. Particularly size effects persist over time. Their overall conclusion is that rescaling local government through municipal reform seems to improve economic steering capacities and fiscal outcomes.

In another Danish study [25] the authors conclude on the basis of survey data collected before and after the Danish municipality reform that when the size of municipalities increases, internal political efficacy drops. We think these findings are questionable. First, the survey data of the opinions of citizens concerning the municipal reform (in 2007) were collected practically immediately after the reform in 2008 (and partly even in the same year of the reform 2007). Municipal reform often evokes strong emotions of lost identity and feelings of resentment and it may take several years before citizens become aware of the benefits. Moreover, decrease of political efficacy completely conflicts with the fundaments of urban scaling: the disproportional increase of social, economic and cultural connections as a function of size. In a study on the municipal reform in New Zealand [26] no evidence was found that municipal reform did improve improvement highway maintenance. Although their analysis is restricted to a just a very specific part of the efficiency in government operations, these authors used their findings to claim that there are no economies of scale. Also in a Swiss study [27] the authors conclude claim that there are no economies of scale in merged municipalities but also here this analysis concerns only the cost side.

On the basis of the analysis of the expenditures of merged municipalities Finnish authors [28] conclude that the per capita expenditure increased more in the merged municipalities than in the comparison group of not-merged municipalities. Here we can notice that urban scaling means a disproportionally increase of the gross urban product and thus it is understandable that the merged municipalities have more to spend. In an Israeli study the author finds evidence for efficiency gains arising from municipal reform in his country [29]. He finds that municipal reform resulted in lower levels of expenditures and has not seemed to decrease the quality of services. Although the empirical work is restricted to relatively small municipalities, this author suggests that as municipalities' size increases the benefits arising from a municipal reform decrease. Apart from the fact that only the cost side and not the benefits are looked at, we also found no indications whatsoever in the literature that a clear distinction is made in types of municipal reform, particularly the merging of suburban municipalities with the central city in a compact and densely connected urban area, versus the merging of more rural and less directly connected municipalities.

## 3. Germany

### 3.1 Data

Our second case is Germany. We now deal with a country about fifteen times larger in population size as compared to Denmark. Germany with about 82 million inhabitants consists of sixteen federal states, the *Bundesländer*. In total, Germany currently has 106 kreisfreie Städte (cities of which the surrounding urban region belongs to the municipalities of the city; together a population of about 27,000,000) and 296 Kreise (regions around smaller cities consisting of several municipalities[6]; together about 55,000,000 population). We analyzed the scaling of these kreisfreie cities and Kreise for the different Bundesländer. We clustered the Bundesländer into five regions: Nordrhein-Westfalen (North Rhine-Westphalia) (western part of Germany); Baden-Württemberg and Bayern (Bavaria) (southern part of Germany); Hessen (Hesse), Rheinland-Pfalz (Rhineland-Palatinate), and Saarland (middle part of Germany); Bremen, Hamburg, Niedersachsen (Lower Saxony), and Schleswig-Holstein (northern part of Germany); Berlin, Brandenburg, Mecklenburg-Vorpommern, Sachsen (Saxony), Sachsen-Anhalt (Saxony-Anhalt) and Thüringen (Thuringia) (eastern part of Germany).

We collected the gross urban product, number of jobs in different business sectors and population of all the kreisfreie cities and the Kreise; we focus in this report on the gross urban project (GUP). As in the Danish case, we collected for all the kreisfreie cities and Kreise the (land) surface areas[7]. Moreover, we idenified for all 238 Kreise in the western part of Germany the municipalities belonging to a Kreis and collected for these nearly 9,000 municipalities data on population and surface area.

---

[6] The data on gross domestic product and population at the level of kreisfreie Städte and Kreise were provided by the Statistisches Bundesamt https://www.destatis.de, data in www.vgrdl.de. Data are compiled and analyzed in file D-UrbScal.xlsx.

## 3.2 Scaling Analysis

Also for Germany we find that the gross urban product scales superlinearly with population for both the kreisfreie cities as well as for the Kreise. But we see remarkable differences between regions. In Fig 8 the results are shown for the traditionally industrial western part of Germany, Nordrhein-Westfalen. Largest city is Cologne, with a population of about 1,100,000. The kreisfreie cities scale with 1.33, and the Kreise with 1.07. Generally, Kreise underperform as compared to the kreisfreie cities. We see that two cities, Düsseldorf and Bonn, take a strikingly strong position. This illustrates the problematic question when measuring points should be considered as outliers. Leaving Düsseldorf out of the analysis, the exponent becomes 1.27. Leaving both Düsseldorf and Bonn out of the analysis we find a scaling exponent of 1.25. This shows as in the case Denmark how the scaling exponent depends on the positions of one or a few cities in the entire set of cities. Nevertheless, the above also shows that the measuring scaling exponent does not change dramatically and remains, for North Rhine-Westphalia, relatively high.

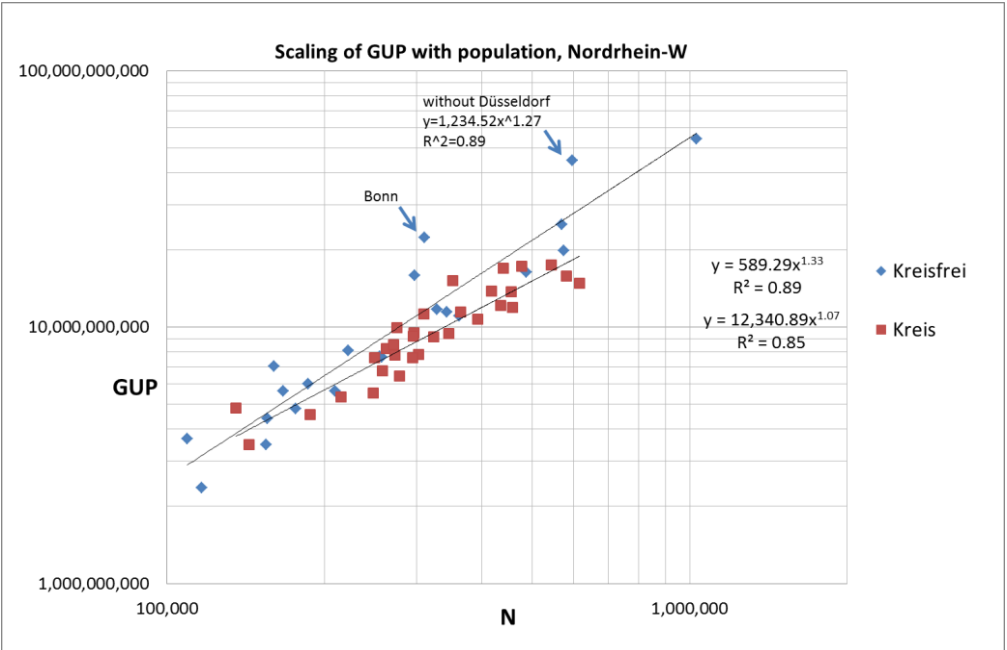

*Fig 8. Scaling of GUP (in €) for kreisfreie cities and Kreise in Nordrhein-Westfalen (data: average 2012-2014).*

Fig 9 shows the results of the economically most booming southern part of Germany, the states Baden-Württemberg and Bayern (Bavaria). Largest city in this region is Munich with a population of about 1,500,000. The kreisfreie cities scale with 1.08, and the Kreise with 1.17. Again the Kreise underperform as compared to the kreisfreie cities, and we will see this also for all other regions discussed hereafter. Also in this analysis we see city with an extraordinary position: Ingolstadt. The GUP of this city is relatively very high because Ingolstadt (about 140,000 inhabitants) is home to the headquarters of the automobile manufacturer Audi and the headquarters of the electronic stores Media Markt and Saturn. Leaving Ingolstadt out of the analysis, the measured exponent does not change much, see Fig 9. Also two other large car industries have their headquarters in this part of Germany: Mercedes-Benz in Stuttgart and BMW in Munich. We see also a Kreis with a very strong position: Landkreis Munich. This Kreis is the region surrounding the city of Munich and benefits greatly from the booming economy of Munich. If we leave this Kreis out of the analysis, the exponent becomes 1.13.

---

[7] For instance, see https://en.wikipedia.org/wiki/Frankfurt.

What could be the cause of the difference in scaling exponent of the cities between North Rhine-Westphalia on the one hand and Baden-Württemberg and Bavaria on other? We think that the most plausible explanation is that in North Rhine-Westphalia the two largest cities, Cologne and as we discussed above particularly Düsseldorf, have a high GUP as compared to most other cities which are considerably less flourishing (with exception for Bonn and also Münster). In such cases the regression line becomes steeper and hence we find a larger exponent. In contrast to this situation, most of the cities in Baden-Württemberg and Bavaria, not only the largest ones, are economically booming. As a result the regression is less steep and we have a lower exponent.

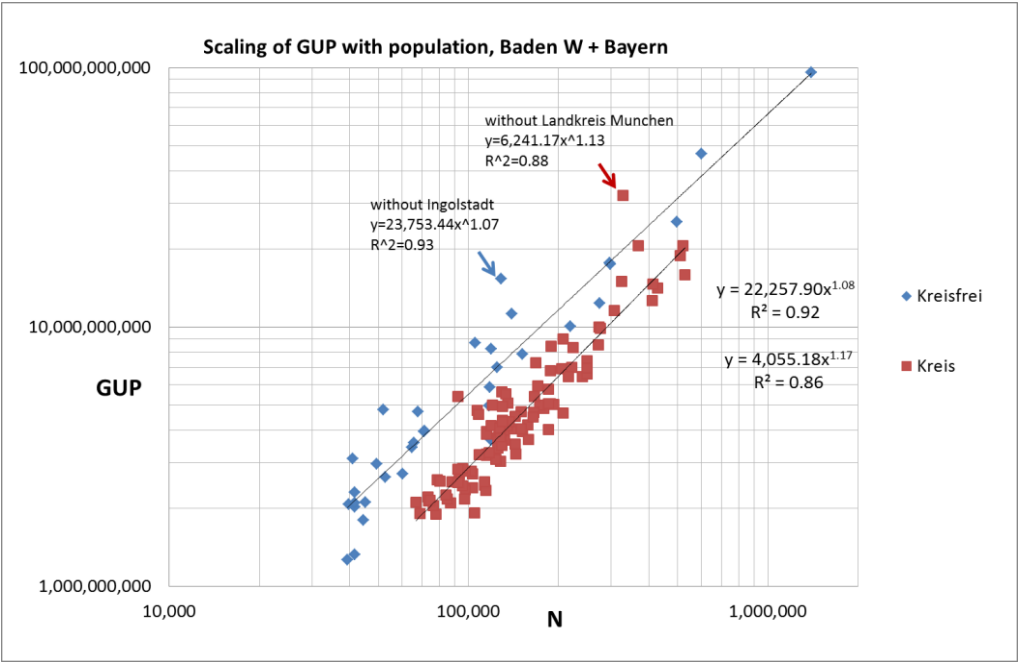

*Fig 9. Scaling of GUP (in €) for kreisfreie cities and Kreise in Baden-Württemberg and Bayern (data: average 2012-2014).*

In Fig 10 the results for the middle part of Germany, the States Hessen (Hesse), Rheinland-Pfalz (Rhineland-Palatinate), and Saarland are presented. Largest city in this region is Frankfurt with about 750,000 inhabitants. The kreisfreie cities scale with 1.31, and the Kreise with 1.28.

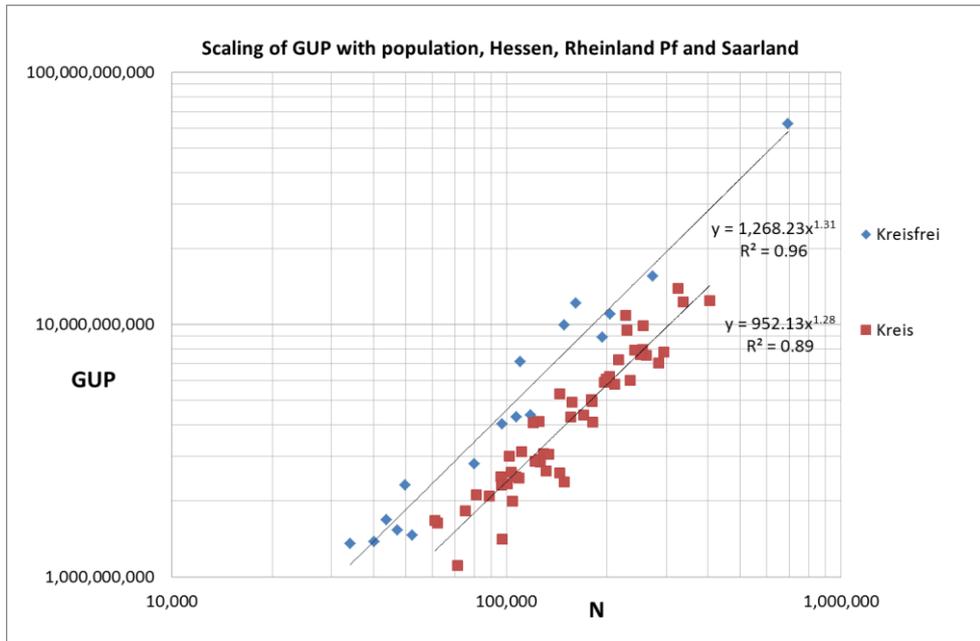

*Fig 10. Scaling of GUP (in €) for kreisfreie cities and Kreise in Hessen, Rheinland-Pfalz and Saarland (data: average 2012-2004).*

Fig 11 shows the results for the northern part of Germany, i.e., the states of Bremen, Hamburg, Niedersachsen (Lower Saxony), and Schleswig-Holstein. Largest city in this region is Hamburg with about 1,800,000 inhabitants. The kreisfreie cities scale with 1.09, and the Kreise with 1.14. In this figure we see a city which clearly is an outlier. This city is Wolfsburg (about 125,000 inhabitants) with a very high GUP because it is the location the Volkswagen headquarters and the world's biggest car plant, production of 815,000 cars per year (2015), 70,000 employees in Wolfsburg alone. Measured in GUP per capita, Wolfsburg is one of the richest cities in Germany. Leaving Wolfsburg out of the analysis we find a scaling exponent 1.11.

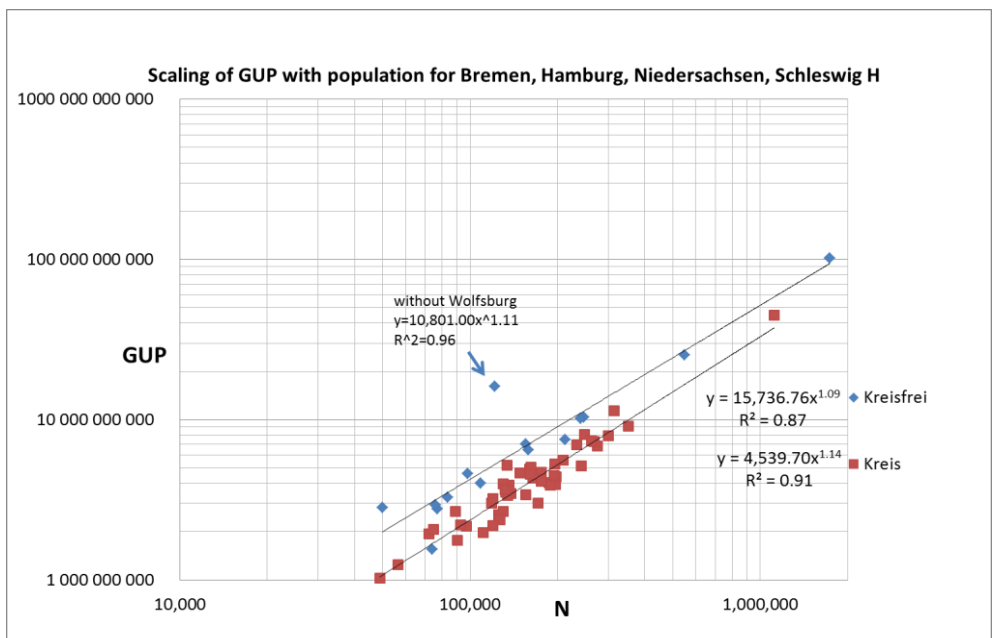

*Fig 11. Scaling of GUP (in €) for kreisfreie cities and Kreise in Bremen, Hamburg, Niedersachsen, and Schleswig-Holstein (data: average 2012-2014).*

In Fig 12 the results are shown for the economically problematic eastern part of Germany, the states Berlin, Brandenburg, Mecklenburg-Vorpommern, Sachsen (Saxony), Sachsen-Anhalt (Saxony-Anhalt) and Thüringen (Thuringia). Largest city in this region is Berlin with a population of about 3,700,000. We see that in this region urban scaling is hardly significant: the kreisfreie cities scale with 1.03, and the Kreise with 1.02. In all other regions of Germany the kreisfreie cities show scaling with exponent between 1.08 and 1.33 and for the Kreise between 1.07 and 1.28. Apparently the mechanisms behind scaling, particularly the size-based non-linear reinforcing of the socioeconomic links in networked systems, hardly work in this part of Germany with its difficult economic development. A possible explanation is the decline in population in the eastern part of Germany and particularly the move away of talented people to other regions in Germany. Without Berlin the scaling exponent is 1.06 which is an indication that Berlin is economically underperforming.

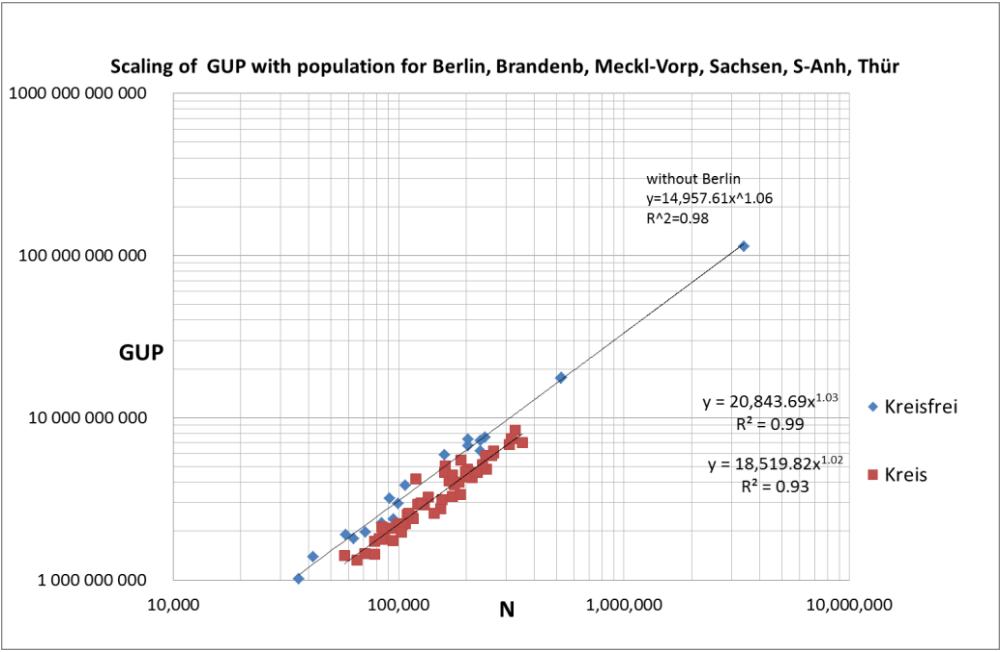

*Fig 12. Scaling of GUP (in €) for kreisfreie cities and Kreise in Berlin, Brandenburg, Mecklenburg-Vorpommern, Sachsen, Sachsen-Anhalt, and Thüringen (data: average 2012-2014).*

In Table 1 we present an overview of the scaling exponents for the kreisfreie cities and Kreise in the five regions. To illustrate the economic East-West-South division of Germany we show in Table 2 for the five regions the gross domestic product per capita in absolute terms and relative compared to Germany as a whole.

| *Region* | *Krfr cities* | *Kreise* |
|---|---|---|
| western | 1.33 | 1.07 |
| southern | 1.08 | 1.17 |
| middle | 1.31 | 1.28 |
| northern | 1.09 | 1.14 |
| eastern | 1.03 | 1.02 |

*Table 1. Scaling exponents for the kreisfreie cities and Kreise in the five German regions.*

| Region | GDP/capita in € | rel. GDP/capita |
|---|---|---|
| western | 35,947 | 1.00 |
| southern | 41,334 | 1.15 |
| middle | 37,515 | 1.04 |
| northern | 35,768 | 0.99 |
| eastern | 27,414 | 0.76 |
| | | |
| **Germany** | **36,003** | **1.00** |

*Table 2. Gross domestic product per capita of the five regions in absolute terms and relative compared to Germany as a whole (data: average 2012-2014).*

We see the wealthy position of the southern region in contrast to the substantially less wealthy situation in the eastern part of Germany. The difference in GDP per capita is a factor 1.15/0.76=1.64. In Fig 13 we combine the results for the kreisfreie cities of all regions. Again clearly visible is the difference in wealth between, for instance, the southern part of Germany (red squares, Baden-Württemberg and Bayern) and the eastern part (blue squares, Berlin, Brandenburg, Mecklenburg-Vorpommern, Sachsen, Sachsen-Anhalt, and Thüringen). By taking the scaling results for only the southern and the eastern part of Germany this difference shows up even more strikingly, see Fig 14.

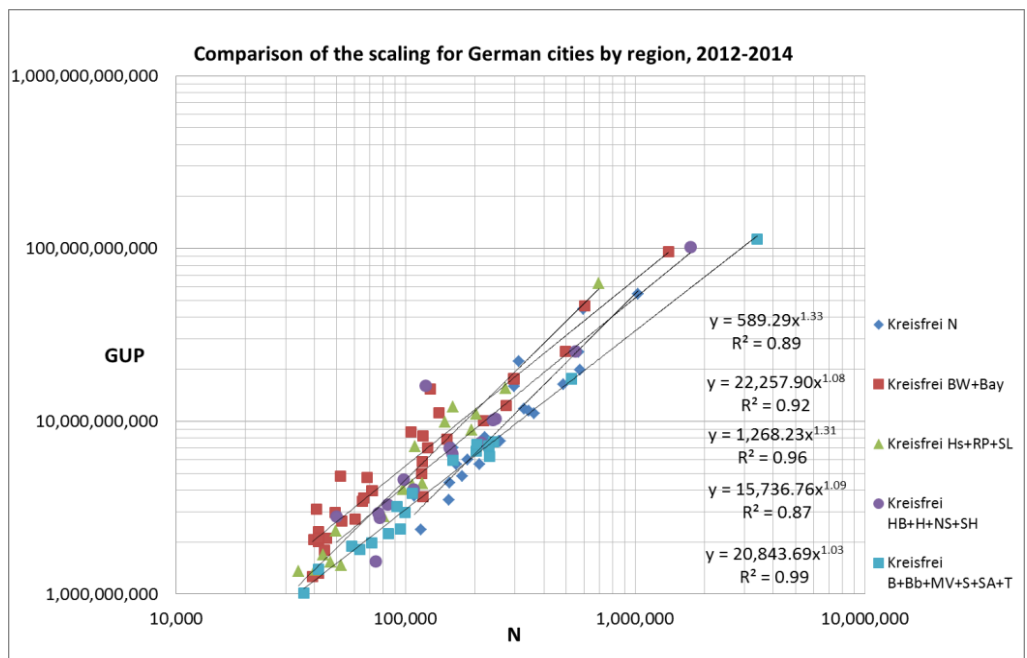

*Fig 13. Comparison of the scaling of GUP (in €) for all kreisfreie cities, grouped by region (compilation of the same data as presented in Figs 8-12).*

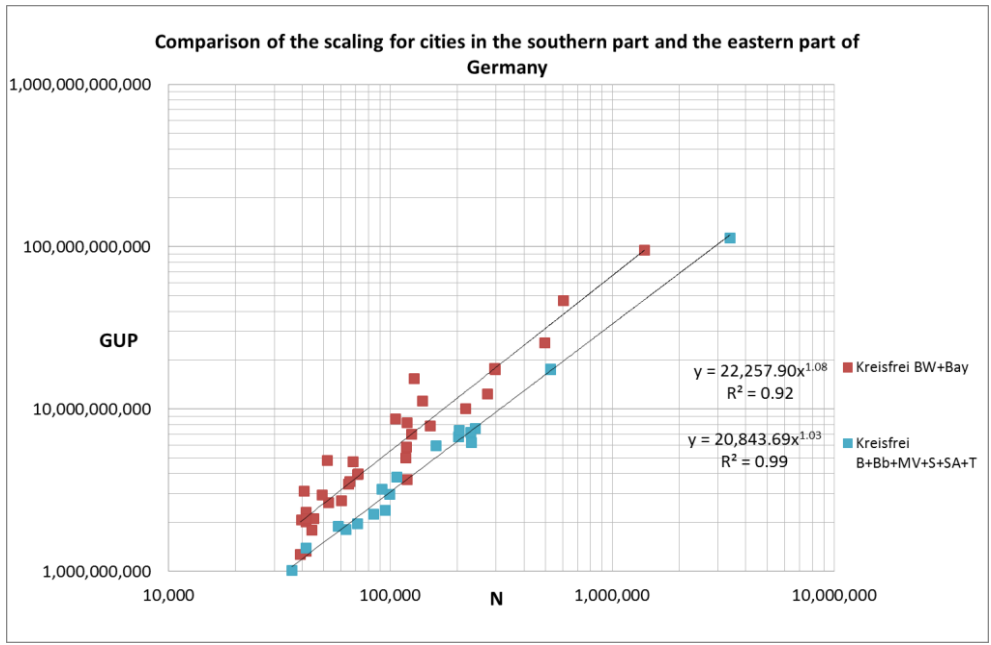

*Fig 14. Same as Fig 13, but now only the southern region in comparison with the eastern region.*

Using the parameters of the measured scaling, we find a difference in average GUP of about 1.96, somewhat higher than the difference in the GDP per capita as calculated on the basis of Table 2. The above findings clearly show a problem in the measurement of scaling in an entire country: the numerous smaller kreisfreie cities in the southern part are at a high level of GUP, thus they will 'lift' the regression line at the lower population side thereby lowering considerably the exponent of the scaling. Fig 15 shows this effect: if we calculate the scaling exponent for all German kreisfreie cities together, we find 1.03, this is lower than the scaling exponents of the kreisfreie cities of all German regions, and it is hardly significant. Without the largest city, Berlin, the scaling exponent is 1.05.

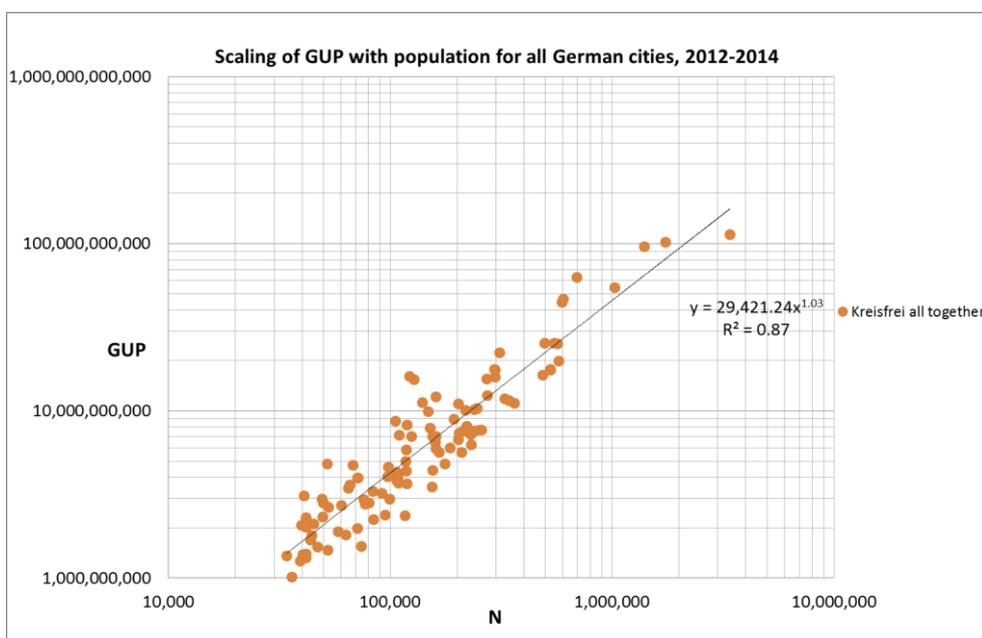

*Fig 15. Scaling of GUP (in €) with population of all kreisfreie cities together.*

A crucial element in our study is the question how governance structures influence socioeconomic performance. We made the following analysis to find first indications. In a number of Kreise the administrative centers are cities that are not kreisfreie cities (because they formally belong to a Kreis) but they can be considerably larger than smaller kreisfreie cities[8]. As a consequence, there is an overlap in population size as well as density between kreisfreie cities and Kreise. In such cases, Kreise are densely populated urban regions, just as the kreisfreie cities. We created a set of kreisfreie cities and of Kreise with similar population size and similar density (>400 inhabitants/km$^2$) and calculated the scaling exponents for both groups. The results are shown in Fig 16. We see that the one-municipality urban regions (the kreisfreie cities) overperform the multi-municipality urban regions.

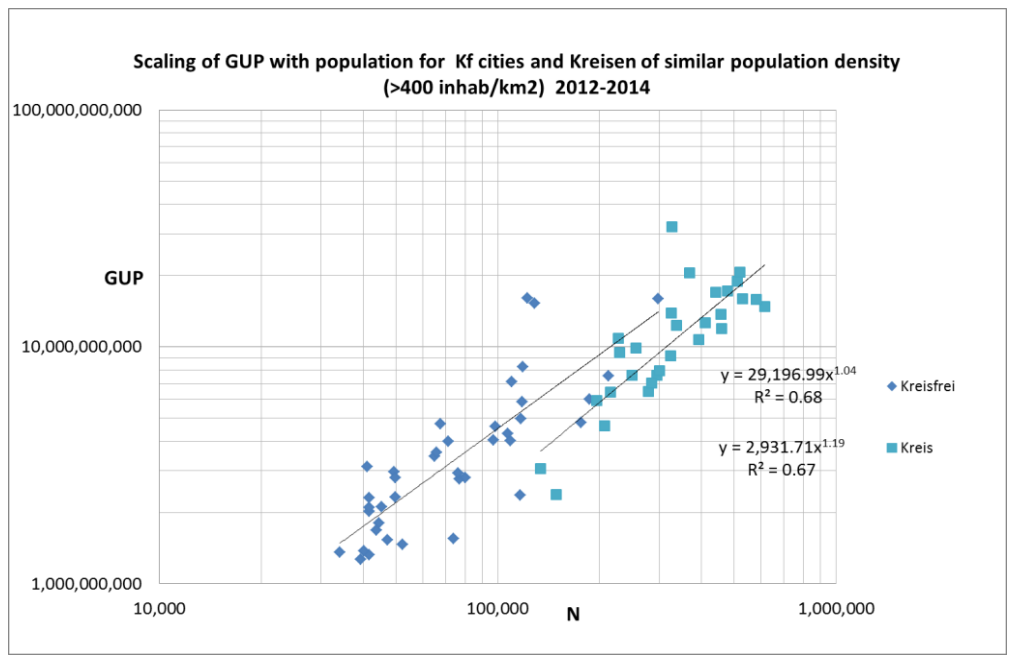

*Fig 16. Comparison of the scaling (GUP, in €) of kreisfreie cities with Kreise of similar population density (Germany as a whole).*

3.3 Residual Analysis and Comparison with Other Socio-Economic Data

We calculated for all German kreisfreie cities and Kreise the residuals of the scaling equations. Residuals are a measure of the deviations of the *observed* (i.e., real) value from the *expected* value as established by the scaling equation. Qualitatively speaking, residuals are a measure of the deviation of a city, municipality, or Kreis from the expected value given by the regression line through all measuring points of a specific set. (see the Appendix for a discussion of the mathematical procedure to calculate the residuals). Analysis of the residuals may reveal local characteristics of individual cities in terms of success or failure relative to other cities. Positive residuals indicate that a city performs better than expected. We did not find any significant relation between population density of both the kreisfreie cities as well as the Kreise and the value of the residuals.

---

[8] An example is Neuss (Nord Rhine-Westphalia) with about 155,000 inhabitants, but this city is not kreisfrei, it is the administrative center of the Rhein-Kreis Neuss which has a population of about 450,000. The Bavarian city Schwabach on the other hand with about 41,000 inhabitants is a kreisfreie city.

The intriguing question now is: what about the relation between the *residuals* calculated on the basis of GUP scaling with population on the one hand, and on the other hand the socio-economic position of a city as assessed by a combination of a large number of different quantitative indicators? The German socioeconomic research agency Prognos AG[9] performed an investigation of the future perspectives of (kreisfreie) cities and regions (Kreise) in Germany on the basis of a number of indicators and published the results in the report *Zukunftatlas 2016* [30]. In total 29 indicators related to demographics, job market, competition and innovation, welfare and social life quality were used to assess strength and dynamism of cities and regions. On the basis of these assessments, a ranking of all cities and regions was created. For details we refer to the above mentioned report [30].

Fig 17 illustrates the economic differences between the regions in Germany based on perceived opportunities for the near future. We clearly see the division between the eastern region and the other regions in Germany. By selecting the kreisfreie cities from the Prognos ranking, we find that of the *top-20 cities* 17 have a residual larger than 0.15, see Table 3, upper part. For the *bottom-20 cities* we find that 16 have a residual smaller than -0,15, see Table 3 lower part.

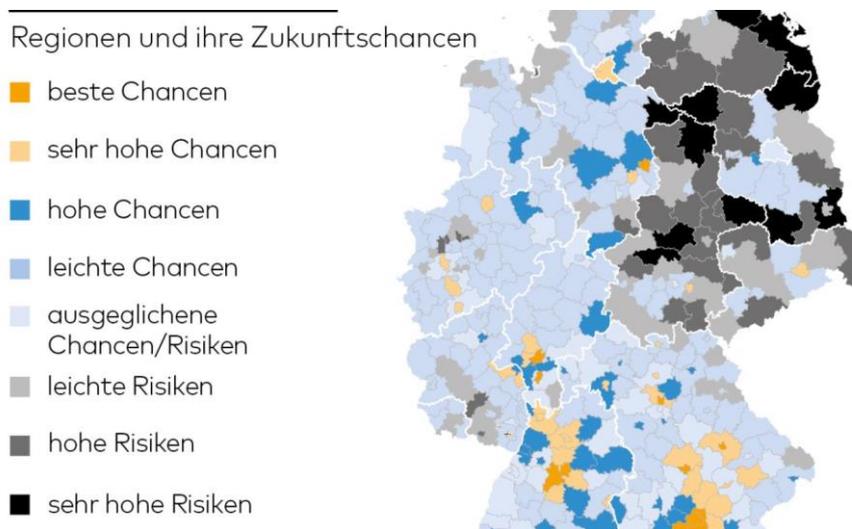

*Fig 17. Assessment of future economic opportunities for the German regions.*[10]

---

[9] See www.prognos.com.
[10] *https://www.welt.de/wirtschaft/article155737236/Diese-zehn-Regionen-haben-die-besten-Zukunftschancen.html*.

| Prognos ranking | residual | kreisfreie city | | Prognos ranking | residual | kreisfreie city |
|---|---|---|---|---|---|---|
| 1 | 0.42 | München | | 87 | -0.21 | Remscheid |
| 2 | 1.05 | Ingolstadt | | 88 | -0.68 | Delmenhorst |
| 3 | 1.15 | Wolfsburg | | 89 | -0.18 | Lübeck |
| 4 | 0.68 | Erlangen | | 90 | -0.46 | Halle |
| 5 | 0.57 | Stuttgart | | 91 | -0.36 | Suhl |
| 6 | 0.46 | Darmstadt | | *92* | -0.15 | Krefeld |
| 7 | 0.71 | Frankfurt aM | | 93 | -0.51 | Gera |
| 8 | 0.65 | Regensburg | | 94 | -0.27 | Hagen |
| 9 | 0.21 | Heidelberg | | *95* | *-0.07* | *Wilhelmshaven* |
| 10 | 0.50 | Ulm | | 96 | -0.33 | Cottbus |
| 11 | 0.26 | Hamburg | | 97 | -0.17 | Schwerin |
| 12 | 0.53 | Düsseldorf | | *98* | *-0.14* | *Neumünster* |
| 13 | 0.63 | Coburg | | 99 | -0.26 | Duisburg |
| *14* | *-0.27* | *Dresden* | | 100 | -0.46 | Oberhausen |
| 15 | 0.28 | Landshut | | 101 | -0.17 | Pirmasens |
| 16 | 0.30 | Würzburg | | 102 | -0.40 | Brandenburg ad H |
| 17 | 0.30 | Bamberg | | 103 | -0.62 | Herne |
| *18* | *-0.16* | *Jena* | | 104 | -0.44 | Dessau-Roßlau |
| *19* | *-0.01* | *Braunschweig* | | *105* | *-0.12* | *Bremerhaven* |
| 20 | 0.51 | Bonn | | 106 | -0.36 | Gelsenkirchen |

*Table 3. Left part: Top-20 cities; right part: bottom-20 cities. Cities in italics have residuals lower than 0.15 in the case of the top-20, and higher than -0.15 in the case of the bottom-20.*

We see that there is a strong relation between the measured residuals in this study and the assessment of future perspectives of cities by the Prognos method.

### 3.4 Urban Scaling and Mono- versus Polycentrality

In our analysis of urban scaling in Denmark we discussed the scaling behavior for specific subsets within the total set of all municipalities. One of the distinctions we made was related to centrality in terms of density within a municipality. With centrality we measure the extent to which residents, or jobs, or GUP, are divided over all cities/towns within a municipality. Given the availability of data, the most workable definition of centrality is the ratio of the population of the main city/town in the municipality to the total population of the municipality. As we discussed in the case of Denmark, in municipalities around larger cities at least half of the total population lives in the central city. Thus, for the typical urban areas the centrality will be above 0.50. Such urban areas are therefore mostly *monocentric*. This is typical the situation for the German kreisfreie cities: regions around major cities are one municipality, and by far the most people and jobs are in the central city.

Similar to the centrality of *one specific municipality* (based on the size distribution of cities/towns within a municipality), we can go one aggregation level higher and calculate the centrality of *regions* around smaller cities with generally 10 to 15 municipalities within the region. This is typical the situation for the German Kreise. If the distribution of population over the municipalities within a Kreis is sharply peaked, then one city/town within the Kreis plays the leading role and we have a monocentric region. With a flatter distribution we have a more *polycentric* region. The Zipf-distribution is a *size-rank* distribution of municipalities within a region, for instance within a Kreis, according to their population size. This ranking should theoretically follow a power law with exponent -1.0, but in reality this is certainly not always the case [31]. Deviations from this value can then be used as a centrality measure and a criterion for mono- or polycentricity. For instance, if the power law exponent is -1.5, the distribution is steeper and the Kreis is more monocentric. If the power law exponent is less than -0.4 we have a flat distribution and the Kreis is more polycentric.

Does the distribution of population over municipalities with a Kreis follow a power law? We analyzed the Zipf-distribution of municipalities within all 236 Kreise in the western part of Germany. We find a wide variety in power law exponents and in the statistical significance of these power laws. As examples we show in Fig 18 the Zipf-distribution of 8 Kreise in the German states Baden-Württemberg and Bavaria, based on the largest six municipalities within the Kreis. We notice that in some cases (upper part of Fig 18) there is a power law distribution with high significance, but the exponent varies greatly between –0.43 and -1.18. As discussed earlier, in the case of a flatter distribution (exponent -0.43, Rhein-Neckar Kreis) we have a more polycentric Kreis, and in the case of a steeper distribution (exponent -1.18, Garmisch-Partenkirchen) the Kreis is more monocentric. An example of a more extreme case is Kreis Kelheim, see middle part of Fig 18. This shows the typical problem in the determination of the Zipf exponent: if we include the five largest municipalities of this Kreis in the measurement, we find with high significance an exponent -0.17. But if we include the next municipality the exponent is -0.40, however, as can be expected, with a much lower significance. The lower part of Fig 18 show more of these cases, without fitting the distribution with a power law.

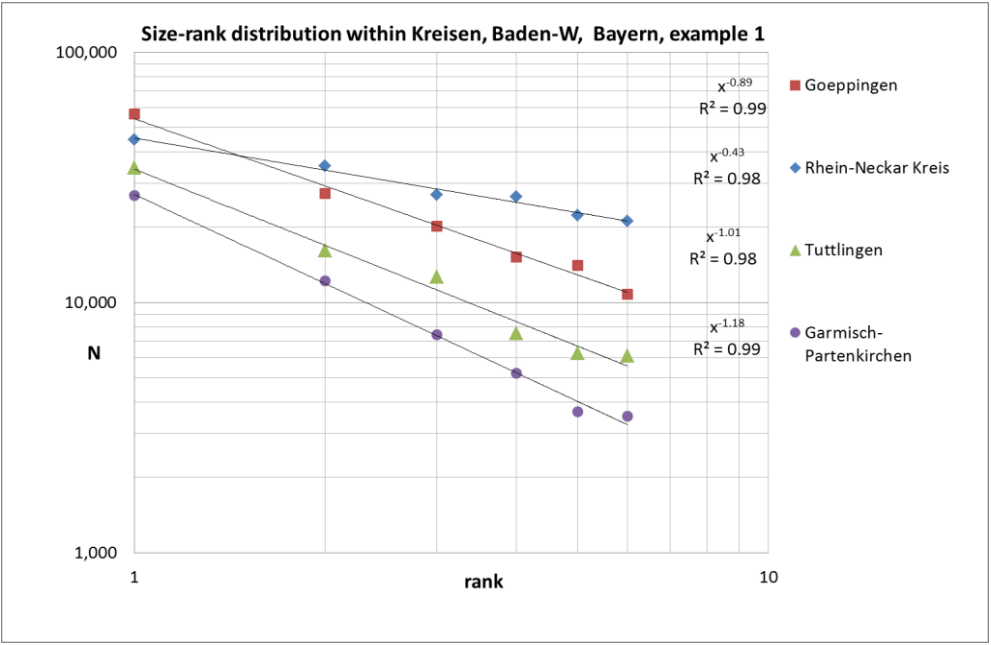

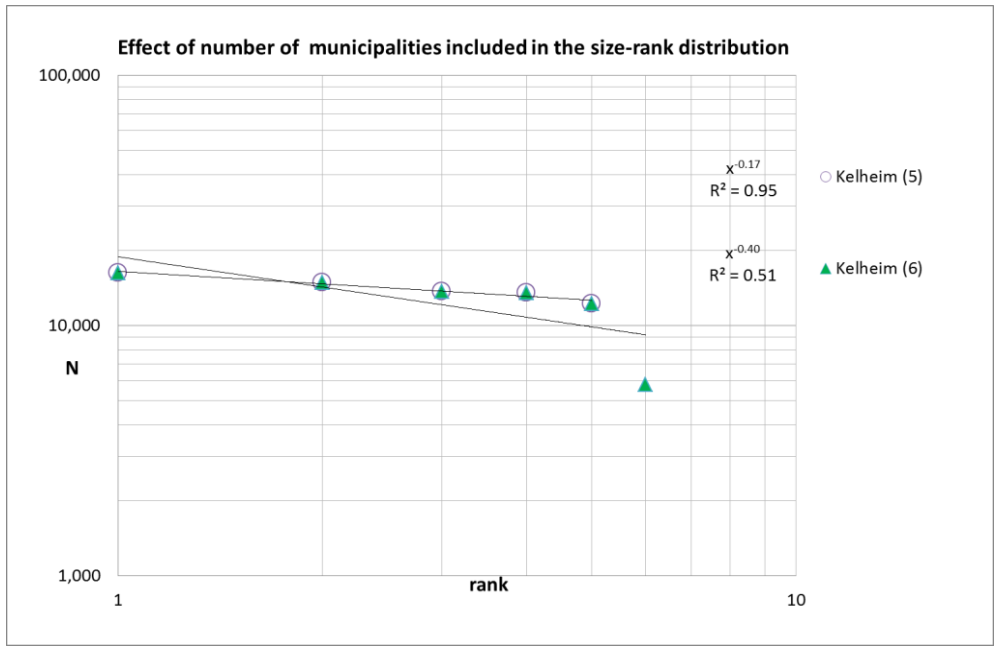

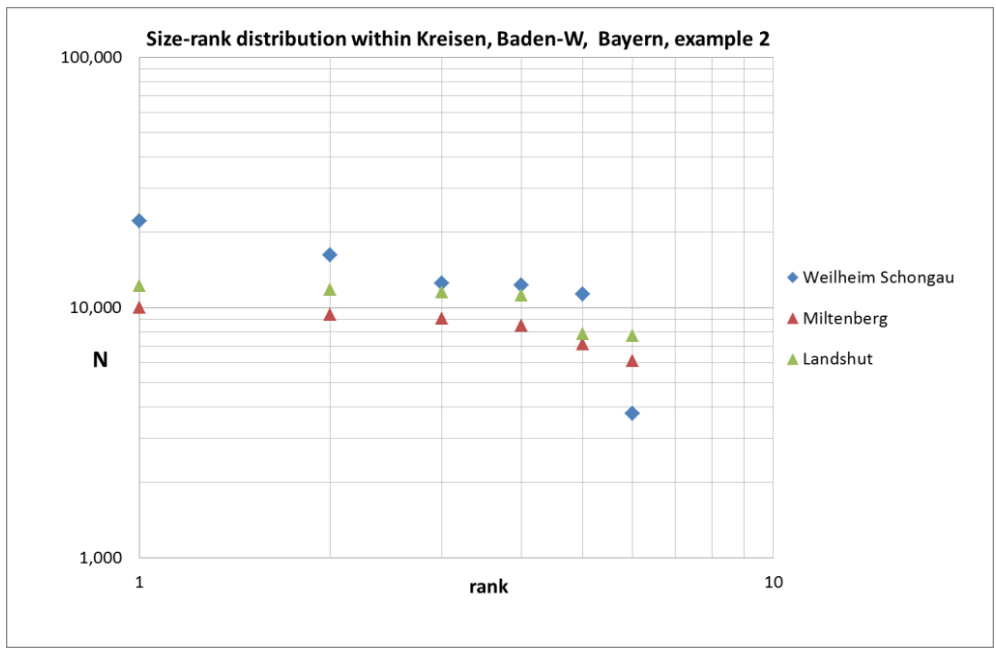

*Fig 18. Examples of size distribution within Kreise in the German states Baden-Württemberg and Bavaria. Upper part: size distribution follows power law; middle part: effect of number of municipalities in the size-ranking distribution; lower part: more examples of this problem.*

Nevertheless we find an exponent with high significance ($R^2 > 0.90$) for the vast majority of Kreise and on this empirical basis we conclude that the calculation of the Zipf-distribution is useful in the analysis of centrality. We also applied two other measures of centrality in which statistical uncertainty plays no role. First a similar one as used in the case of the Danish municipalities but now at the aggregation level of a Kreis, i.e., the ratio of the population of the largest municipality to the total population of the Kreis, denoted as $c_1$; and second the ratio of the population of the largest municipality to the population of the of the second largest municipality in a Kreis, denoted as $c_2$. Comparison of Zipf exponents, $c_1$ and $c_2$ values for all 236 (West-)German Kreise shows that the strongest correlation is between Zipf exponents and $c_1$ values ($R^2$=0.54, so this correlation is not very strong).

We calculated the scaling of GUP with population for all Kreise as a function of the three centrality measures and compare this scaling for the top and the bottom of the centrality distribution. As shown in Fig 19, upper part, we find that the scaling of the Kreise within the top-25% and those within the bottom-25% of the Zipf-exponent show similar exponents but the Kreise in the bottom-25% underperform. The same is the case for centrality measure $c_1$, see middle part of Fig 19. For centrality measure $c_2$ there is no difference between the top and the bottom of the distribution, see lower part of Fig 19. In conclusion we find that the difference in scaling exponent in all these cases is hardly significant, but that the Kreise with a higher Zipf-exponent (in absolute terms) and a higher $c_1$ value (i.e., larger degree of monocentricity) perform better. These differences seem small at first glance when looking at the figures, but given the logarithmic scale, the difference in GUP is quite substantial (about 14% in GUP).

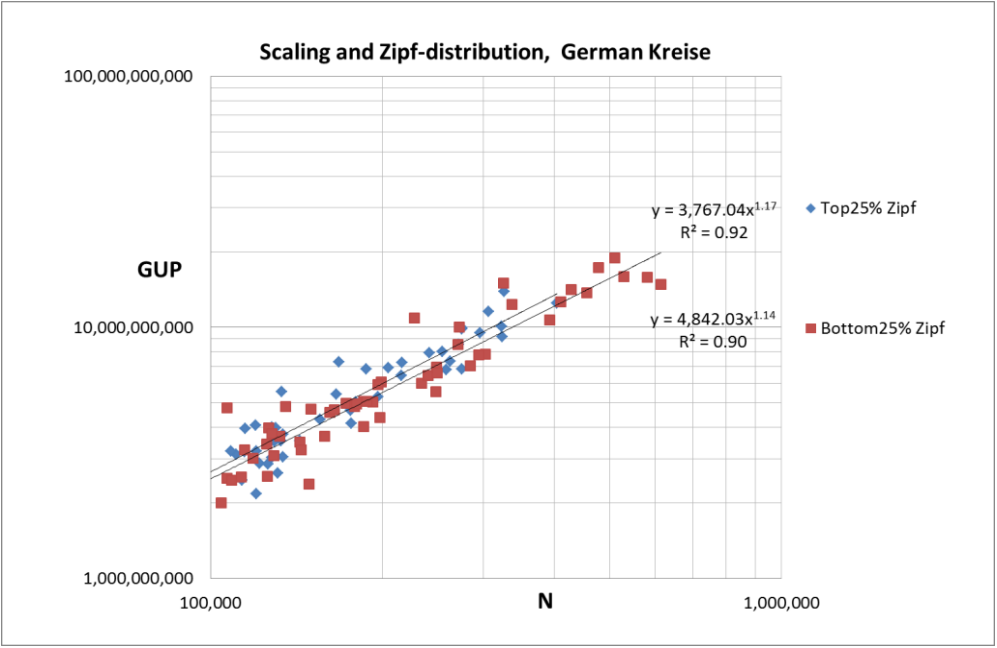

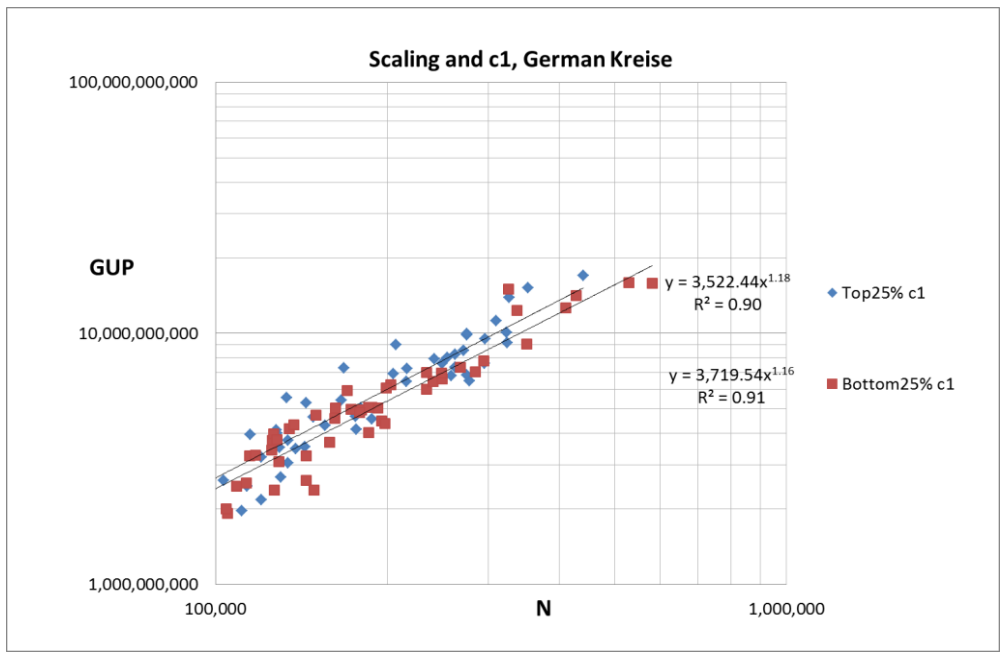

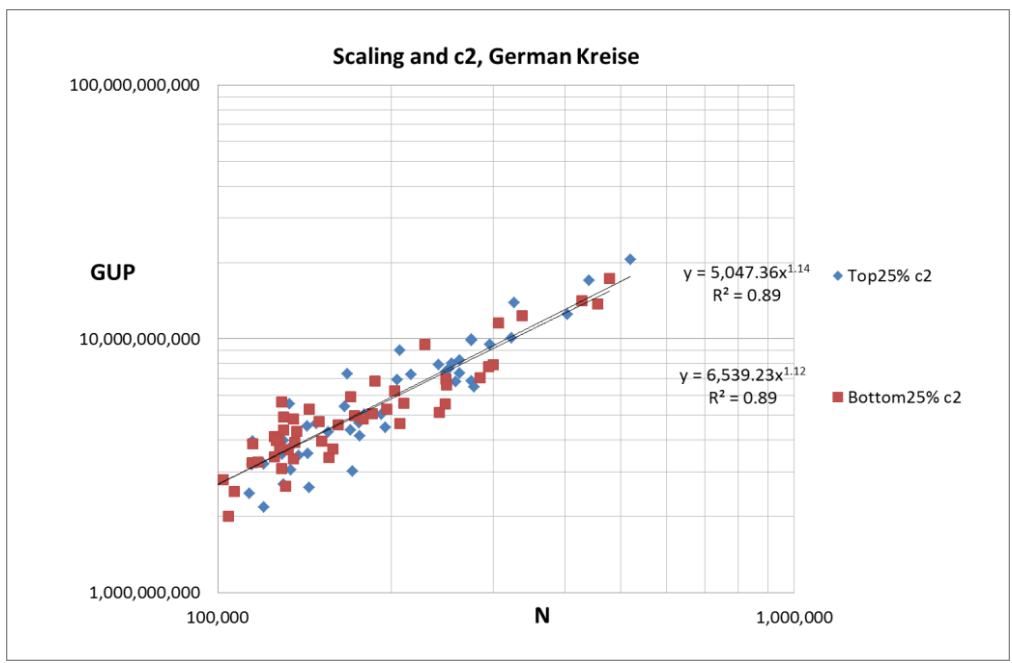

*Fig 19. Scaling of the German (western part) Kreise in the top-25% and the bottom-25% of the Zipf exponent (upper part), $c_1$ values (middle part) and $c_2$ values (lower part).*

## 3.5 Generative versus distributive processes

In Section 2.3 we discussed the issue of generative versus distributive processes related to urban scaling. We showed that an analysis of all Danish municipalities provides evidence that scaling is related to generative processes. Fig 20 shows that we find similar results for Germany.

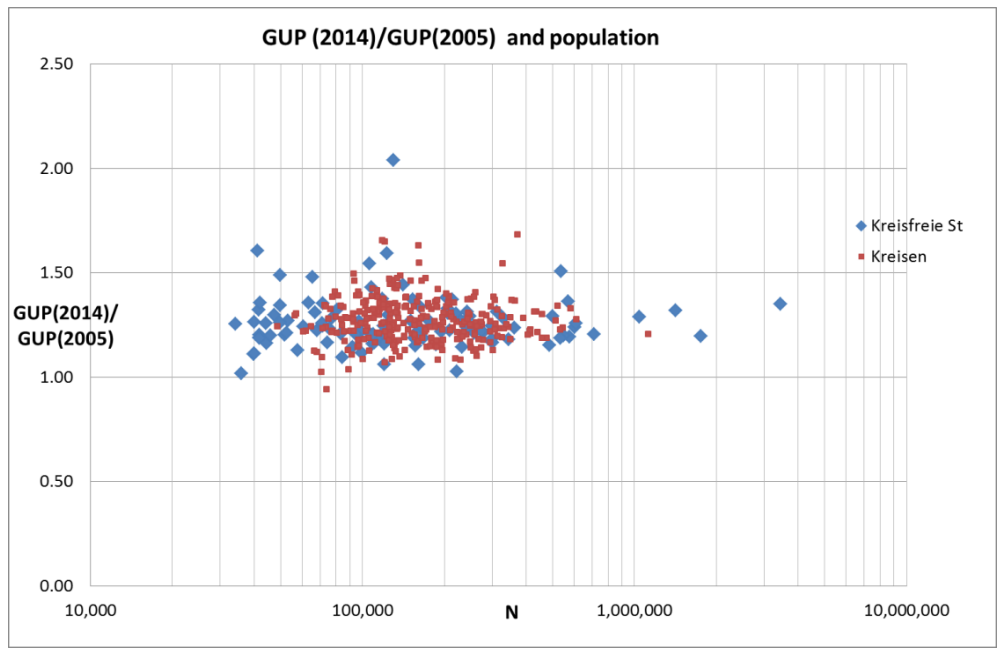

*Fig 20. Increase of GUP between 2005 and 2014 as a function of population size.*

For both the kreisfreie cities as well as the Kreise in Germany no significant dependence of the ratio GUP(2014)/GUP(2005) on population size is found. Again we conclude that scaling is related to generative processes.

## 4. The Netherlands

### 4.1 Data

The Netherlands with around 17,000,000 inhabitants is administratively divided in 12 provinces and 380 municipalities (December 2018). In this study we focus on two cases: (1) 21 major cities in the Netherlands for which the Central Bureau of Statistics (CBS) has defined urban agglomerations and urban areas; in total the urban areas of these 20 major cities comprise 150 municipalities; and (2) the Province of South-Holland (PZH), the most populated province of the Netherlands (density 1,300 inhabitants/km$^2$) with nearly 3,6 million inhabitants, and 52 municipalities (January 2019); the major urban regions in PZH -Rotterdam, The Hague, Leiden, Dordrecht- are also included in case 1. We collected in both cases for the period 2013-2016 for all municipalities the following variables: (1) number of inhabitants (population); (2) employment (number of jobs); (3) gross urban product; (4) productivity[11]. In addition we collected for all municipalities in case 1 the land surface areas (in km$^2$, total surface area corrected for water surface area) and for each of the 129 suburban cities the distances (in km) to the center of the central city.

---

[11] The GUP data are obtained from the national information system on employment LISA, www.lisa.nl, and the data on the population of municipalities from Statline, the data system of the Netherlands Central Bureau of Statistics (CBS).

## 4.2 Major cities in the Netherlands

### 4.2.1 Scaling Analysis

For major cities the Netherlands Central Bureau of Statistics (CBS) defines two types of agglomerations. First, the *urban agglomeration* which is the central city and the immediately connected suburban cities that are separate municipalities. Second, the *urban area* in which in addition to the urban agglomeration all other suburban cities (again separate municipalities) that are closely socio-economically connected to the central city are included. The largest urban area, Amsterdam, counts 1.7 million inhabitants. We conducted the scaling analysis for *three* urban modalities:

(1) the major cities as a municipality (in total 21);

(2) their urban agglomerations (in total the 21 central cities and 44 suburban cities); and

(3) their urban areas (in total the 21 central cities, the 40 suburban cities in the agglomerations, and in addition 89 suburban cities to complete the urban areas).

In Fig 21 (three parts) we present the scaling of the 21 major cities as well as for their agglomerations and urban areas for the number of jobs, productivity, and GUP. Focusing on the GUP scaling we see that the major cities scale with the following exponents: 1.20 for cities as a municipality (i.e., one-governance structure); 1.16 for the urban agglomerations of these cities and 1.17 for the urban areas of these cities (in these two latter cases: multi-governance structure). On the basis of our earlier calculations [14] we estimate that the 95% confidence interval is within +/- 0.05 around the measured scaling exponents. Our observations suggest a slight decrease of the exponent from central cities as municipalities to urban agglomerations and urban areas, is in good agreement with our earlier analysis [14].

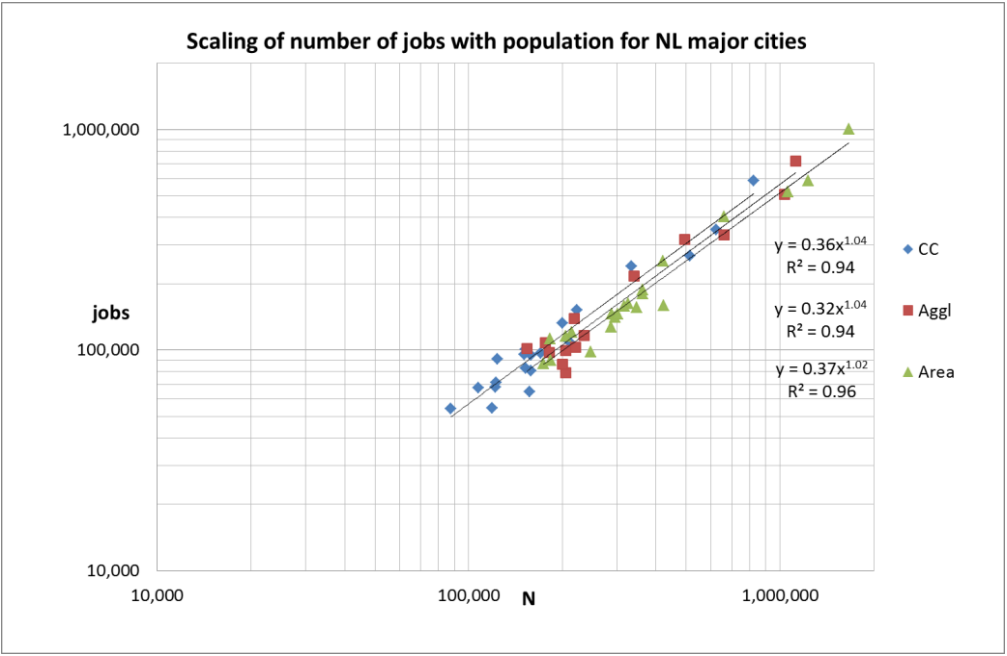

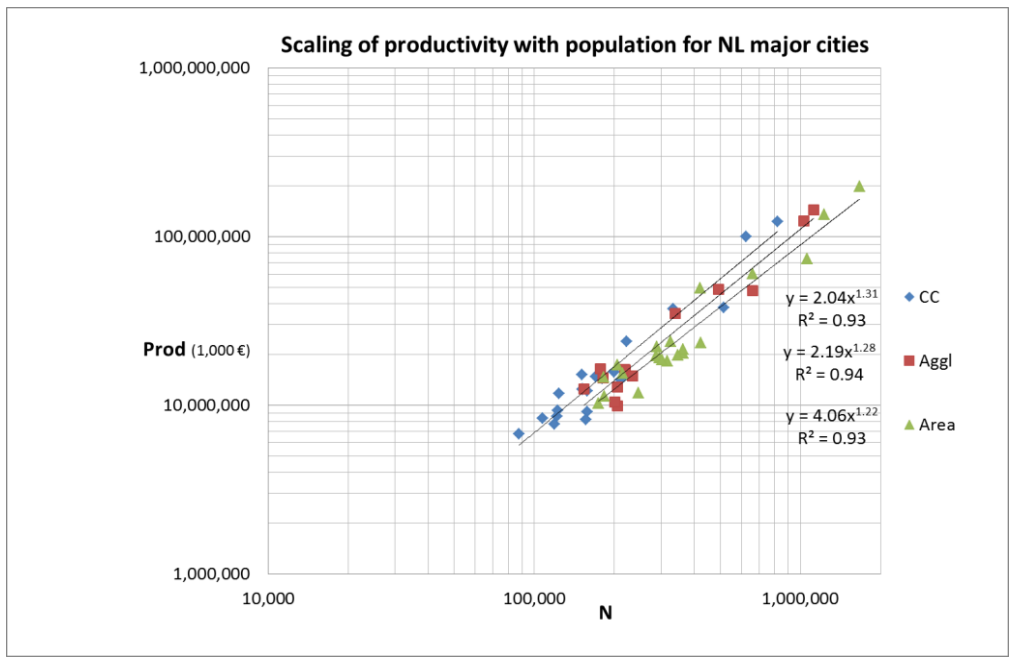

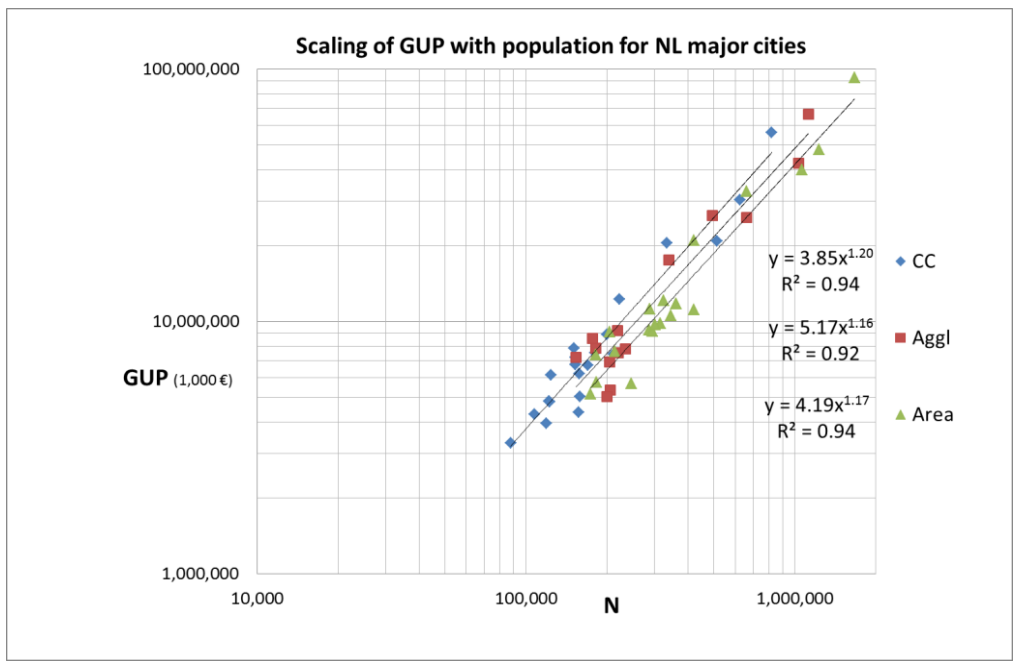

*Fig 21. Scaling of the 21 major cities with number of jobs (first part), productivity (in 1,000 Euro) (second part), and GUP (in 1,000 Euro) (third part). The central cities are indicated with blue diamonds (CC), their agglomeration with red squares (Aggl) and their urban areas with light green triangles (Area).*

Also the remarkable phenomenon discussed in [14] is confirmed again: the absolute value of the gross urban product for both the urban agglomerations and the urban areas is lower than for the central cities as municipalities. Thus, although both types of multi-governance urban region modalities scale with population, they underperform as compared to cities defined as municipalities. In other words, an urban area consisting of one municipality (one-governance) performs substantially better as compared with an urban area with the same number of inhabitants but consisting of several municipalities. And this difference is large in terms of GUP. For instance, the expected value for CC (one-governance) is in the case of 200,000 inhabitants around 20% higher than the expected value for Aggl (multi-governance). This gives a first indication of the profit that can be made by the municipal merging of a central city with its directly connected suburban municipalities. Even if only 10% of these expectations is fulfilled, we are still talking about an amount of 100 million Euros per medium-sized city mainly in the form of thousands of jobs.

In Fig 22 we show the scaling of GUP for all the municipalities in the urban areas, in total 150, which is nearly half of all municipalities in the Netherlands. The most likely explanation of the high scaling exponent 1.27 is that particularly the smaller suburbs are typical residential municipalities and consequently they have a relatively low GUP. This pulls down the regression line resulting in a higher exponent. We see several outliers, i.e., municipalities with an exceptionally high residual. We give two examples. The one at the lower end of the population scale is Zoeterwoude, a municipality in the Leiden agglomeration. This outlier position is mainly due to the presence of the large Heineken beer factory. The outlier at the higher end of the population scale is Haarlemmermeer, a larger municipality in the Amsterdam urban area. Its exceptional position is very well understandable, Amsterdam International Airport Schiphol, the fourth largest airport in Europe, is located in Haarlemmermeer. If Zoeterwoude is removed from the analysis, the scaling exponent is 1.28. Removal of Haarlemmermeer gives a scaling exponent of 1.25.

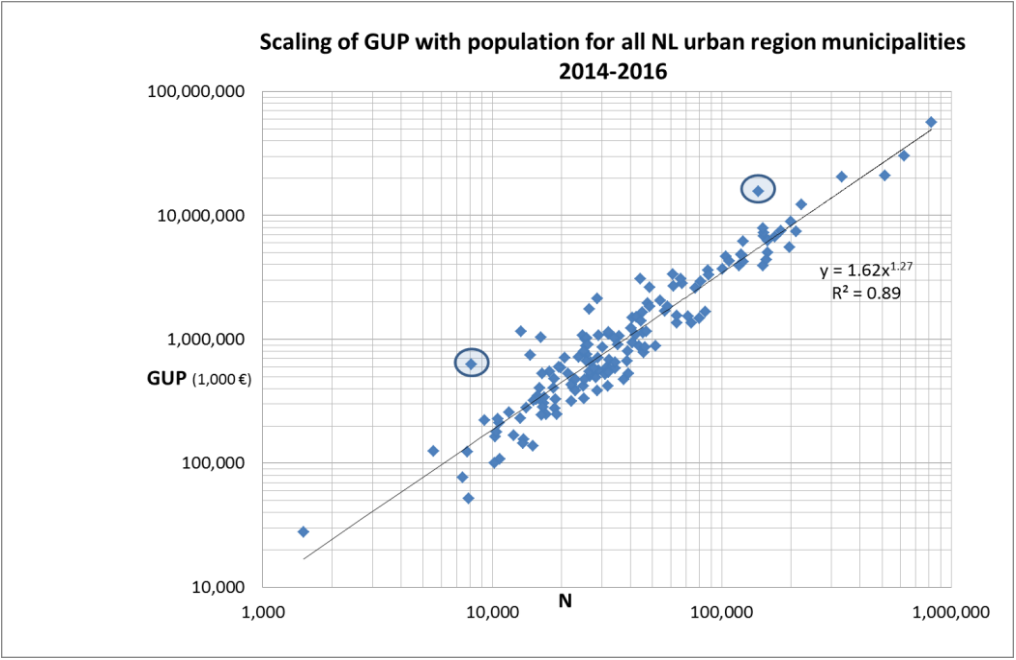

*Fig 22. Scaling of all municipalities in the urban region of the 21 major cities with GUP (in 1,000 Euro). Circles indicate examples of outliers, see discussion in main text.*

*4.2.2 Analysis and Comparison with Other Socio-Economic Data*

We calculated the residuals of the scaling equations given in Fig 21 for the major cities and the suburban municipalities. As discussed earlier in the context of the urban scaling of German cities, analysis of the residuals may reveal local characteristics of individual cities in terms of success or failure relative to other cities [3]. In the Netherlands a socio-economic assessment of the 50 largest cities is published annually in the Atlas voor Gemeenten (AvG, Atlas for Municipalities) [32]. These assessments are predominantly based on qualitative indicators. Of these cities, 32 are within the urban agglomerations/areas of the 21 major cities (either as the central city, or as a major suburb). For these 32 cities we compared their residuals with the ranking position (score) in de socio-economic index of the AvG 2017[12]. These residuals (*resCC*) are related to the scaling of the 21 major cities, i.e., the residuals calculated with the scaling equation of the CC-regression line in Fig 21. To avoid the influence of the individual residuals as well as individual AvG scores, we calculated average values for blocks of five cities according to the ranking of the residuals, see Table 4. For instance, in the second column 0.38 is the average residual of the top-5 cities (in terms of the residuals), and 18.0 is their average AvG score, and so on for the next blocks of five cities.

|  | *resCC* | av res | score | av score |
|---|---|---|---|---|
| Haarlemmermeer | 0.98 |  | 24 |  |
| Utrecht | 0.25 |  | 22 |  |
| Den Bosch | 0.24 |  | 19 |  |
| Zwolle | 0.23 |  | 17 |  |
| Eindhoven | 0.22 | 0.38 | 8 | 18.0 |
| Velsen | 0.18 |  | 12 |  |
| Amsterdam | 0.18 |  | 13 |  |
| Arnhem | 0.14 |  | -11 |  |
| Amstelveen | 0.12 |  | 23 |  |
| Amersfoort | 0.08 | 0.14 | 21 | 11.6 |
| Leeuwarden | 0.04 |  | -3 |  |
| Groningen | 0.03 |  | 2 |  |
| Heerlen | 0.02 |  | -21 |  |
| Leiden | 0.01 |  | 15 |  |
| Maastricht | 0.00 | 0.02 | -9 | -3.2 |
| Hengelo | 0.00 |  | -6 |  |
| Breda | -0.02 |  | 20 |  |
| Delft | -0.03 |  | 0 |  |
| Apeldoorn | -0.05 |  | 9 |  |
| Schiedam | -0.05 | -0.03 | -15 | 1.6 |
| Nijmegen | -0.06 |  | -8 |  |
| Rotterdam | -0.11 |  | -17 |  |
| Zoetermeer | -0.14 |  | 3 |  |
| Dordrecht | -0.16 |  | -19 |  |
| Tilburg | -0.22 | -0.14 | 7 | -6.8 |
| Den Haag | -0.25 |  | -16 |  |
| Enschede | -0.26 |  | -24 |  |
| Haarlem | -0.39 |  | 14 |  |
| Almere | -0.43 |  | -12 |  |
| Vlaardingen | -0.50 |  | -20 |  |
| Leidschendam-Voorburg | -0.65 |  | 4 |  |
| Purmerend | -0.67 | -0.45 | -1 | -7.9 |

*Table 4. Comparison of residuals with scores in the AvG socio-economic review.*

We find a correlation of $R^2=0.79$. This is just a simple test but it confirms the findings in the case of Germany: a significant correlation of the measured residuals based on the GUP scaling with socio-economic indicators from other sources.

---

[12] Ranking data on page 232. To make comparison with the positive and negative residuals easier, we take 25-*a* where *a* is the ranking score given in the AvG.

*4.2.3 Scaling within Urban Areas*

We found that scaling of the municipalities within the Copenhagen agglomeration has a somewhat higher power law exponent (1.24) as compared to the set of all Danish municipalities. Given that the Netherlands has more larger urban areas with suburban municipalities around the central city, we analyzed the urban scaling within the four largest urban areas in the Netherlands, Amsterdam, Rotterdam, The Hague and Utrecht, and within four medium-sized urban areas, Eindhoven, Haarlem, Leiden, Dordrecht [13].

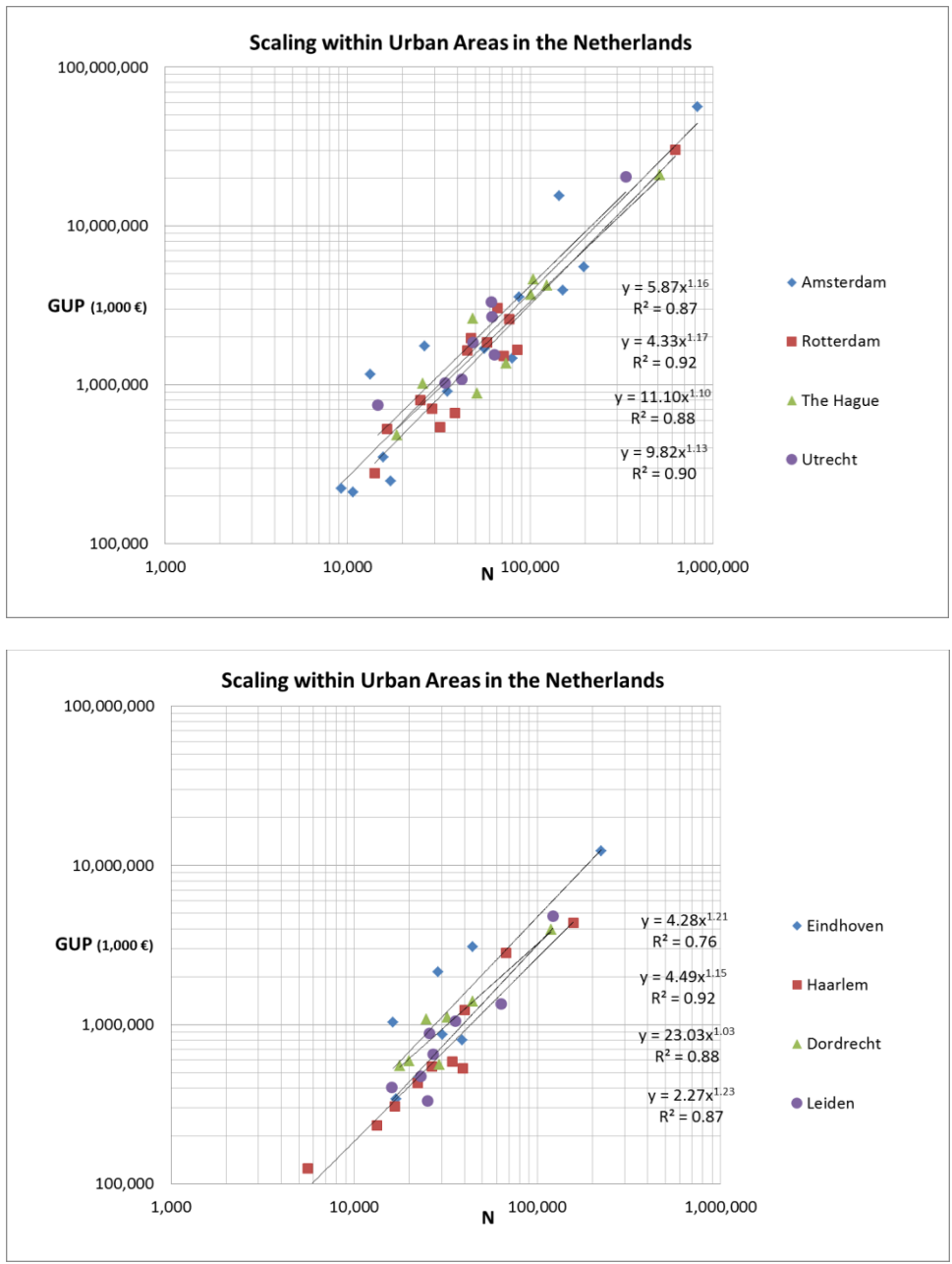

*Fig 23. Scaling of GUP for all municipalities within the urban areas of Amsterdam, Rotterdam, The Hague, Utrecht (upper part of figure), and of Eindhoven, Haarlem, Leiden and Dordrecht (lower part of figure).*

---

[13] The population size of these urban areas are (2018): Amsterdam 1,728,000; Rotterdam 1,256,000; The Hague 1,096,000; Utrecht 682,000; Eindhoven 433,000; Haarlem 425,000; Leiden 350,000; Dordrecht 289,000.

The results of the scaling analysis are presented in Fig 23. We find that the scaling exponents of these urban areas are within a range from 1.03 (Dordrecht) to 1.23 (Leiden) with an average value of 1.15. The results are in line with what we have found for the Copenhagen agglomeration: the scaling within urban areas does not behave differently from urban scaling of, for instance, all cities or municipalities in a country. This, however, does not detract from our findings that multi-governance urban region modalities underperform as compared to one-governance urban areas.

## 4.3 Province of South-Holland

### 4.3.1 Scaling Analysis

In Fig 24 the scaling of the number of jobs (upper part), productivity (middle part) and GUP (lower part) with population for all PZH municipalities is shown. Again we see the municipality of Zoeterwoude as an outlier. As discussed earlier, outliers affect the measured scaling exponent, and we indicate within the figure the scaling exponent without the outlier.

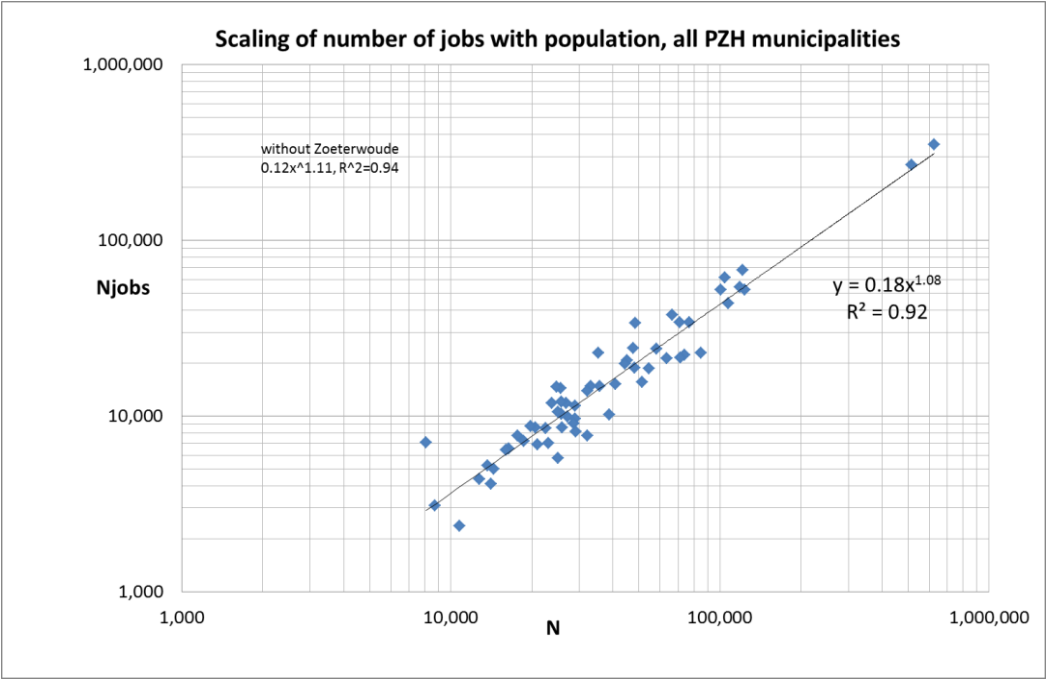

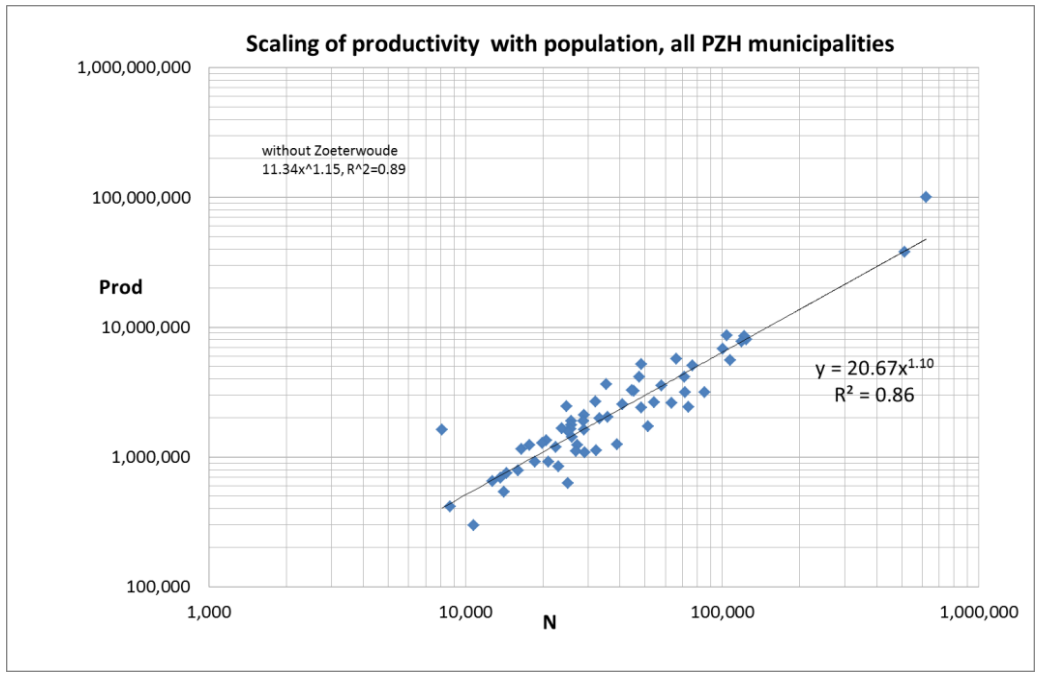

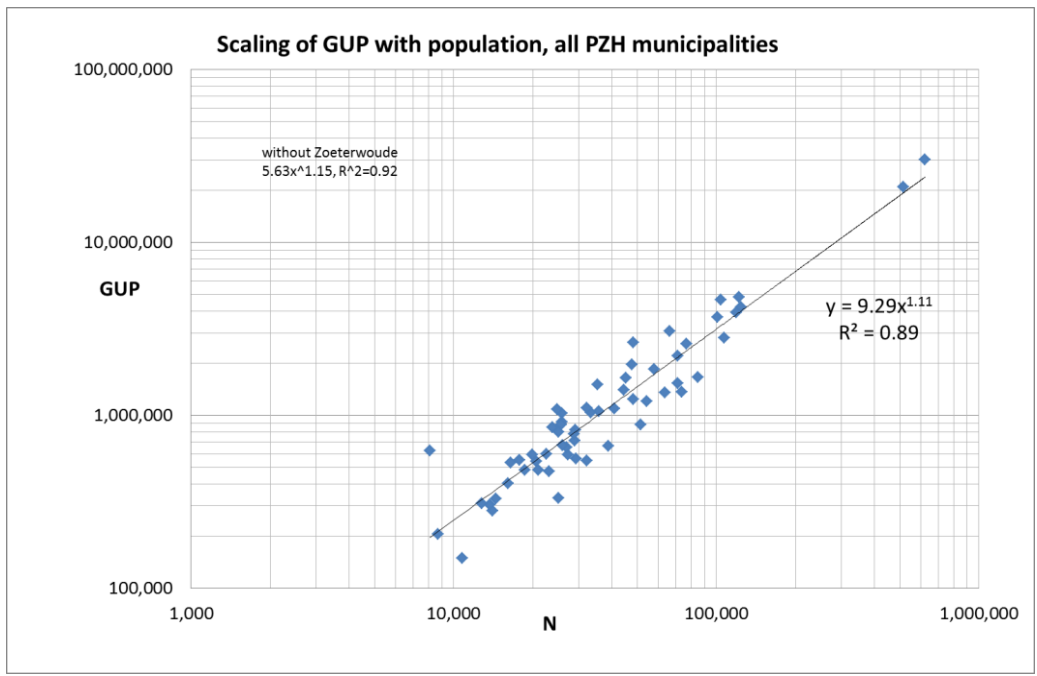

*Fig 24. Scaling of the number of jobs with population (upper part), of productivity (in 1,000 Euro) (middle part) and of GUP (in 1,000 Euro) (lower part) for all municipalities in the Province South-Holland.*

Similar to the analysis of the Danish and German data, we focus on GUP. We find that (without outlier) the entire set of PZH municipalities scales with exponent 1.15, a value again within the range of scaling exponents found in this study.

*4.3.2 Residual Analysis in Provincial and in National Context*

On the basis of the scaling equation found for the GUP (Fig 24, lower part) we calculated the residuals for all PZH municipalities (excluding outlier). The results are shown in Fig 25. We see that of the major cities Rotterdam and Leiden have relatively large positive residuals, whereas the residuals for The Hague ('s-Gravenhage) and Dordrecht are around zero. High positive residuals are found for Rijswijk (The Hague agglomeration) and Sliedrecht (Dordrecht agglomeration). A remarkably low residual is found for Voorschoten (Leiden agglomeration).

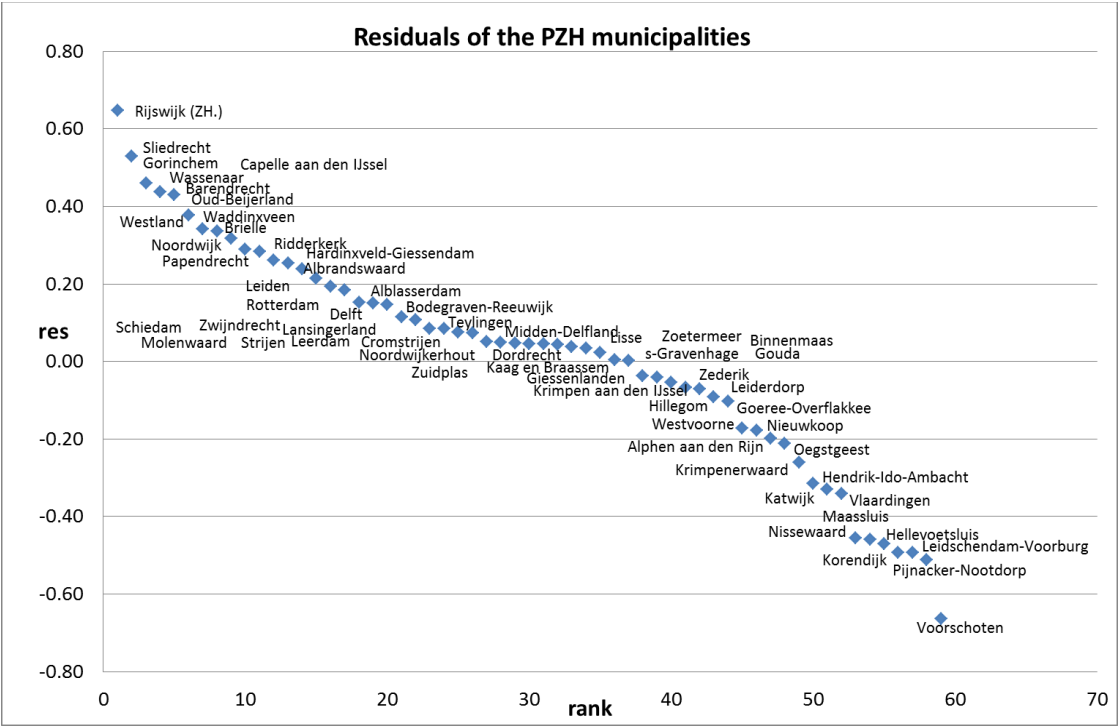

*Fig 25. Residuals of all PZH municipalities[14].*

We now focus on the four major urban regions in PZH (Rotterdam, The Hague, Leiden, Dordrecht). An overview of the population size of these urban regions is presented in Table 5.

|  | *CC* | *U Aggl* | *U Area* |
|---|---|---|---|
| **Rotterdam** | 623,872 | 1,032,380 | 1,230,088 |
| **Den Haag** | 514,596 | 662,605 | 1,061,443 |
| **Leiden** | 121,762 | 204,885 | 345,973 |
| **Dordrecht** | 118,797 | 220,233 | 287,037 |

*Table 5. Population size (average 2014-2016) for the central cities (CC), their agglomerations (U Aggl) and their urban areas (U Area). See Table 6 for the municipalities involved.*

We use the findings of the analysis of the 21 major cities discussed in Section 4.2.1. This means that the PZH urban areas are now assessed in a national context. Table 6 shows the results for the calculation of a set of residuals. The explanation of the table is as follows. In the first column the municipalities within the urban areas are given. Directly below the central city we find the municipalities of the urban agglomeration (for instance in the case of Rotterdam these are the municipalities up to and including Krimpen aan den Ijssel) and in addition the municipalities of the wider urban area (for Rotterdam the municipalities Ridderkerk up to and including Lansingerland). The column *resAll* gives the values of the residuals of the four cities and their suburban municipalities in relation to *the scaling of all municipalities* in the urban areas, in total 150 (shown in Fig 22). In the next column we find *resCC*, the residuals in relation to *the scaling of the 21 major cities*, i.e., the residuals calculated with the scaling equation of the CC-regression line in Fig 21, lower part (these *resCC* are also used in Table 4). It is a measure of *over- or underperformance of cities* (as municipalities) *in a national context*. We see that with the exception of Leiden, the other three cities Rotterdam, The Hague and Dordrecht substantially underperform in this national context. In comparison, the *resCC* values for other major cities in the Netherlands are for instance Amsterdam 0.18, Eindhoven 0.22 and Utrecht 0.25.

---

[14] Excluding Zoeterwoude. Calculated with the same scaling equation as the other PZH municipalities, the residual of Zoeterwoude is 1.27.

Residual *resUAg* is calculated with the scaling equation of the Aggl-regression line in Fig 21, lower part. It is a measure of over- or underperformance of urban agglomerations (central city with the immediately connected suburban cities that are separate municipalities), again in a national context. The results are quite remarkable. The Rotterdam agglomeration does not improve as compared to the Rotterdam *resCC*, which means that currently the suburban municipalities are not in a position to reinforce the Rotterdam urban agglomeration. We observe a similar situation for Leiden. The opposite is found for the Hague and Dordrecht. In our opinion *resUAgCC* is the most interesting measure. If an agglomeration would be a one-governance city, the *expected position* should be on the CC regression line. Thus *resUAgCC* indicates the difference in gross urban product between what the agglomeration would 'earn' if it was a one-governance city, and what it actually earns now, in a fragmented, multi-governance urban system. In all four cases it is clear that the multi-governance structure does not succeed in attaining the expected one-governance level of the gross urban product. Urban agglomerations that perform better in this respect are for instance Amersfoort, Amsterdam, Arnhem, Eindhoven, Groningen, Den Bosch and Utrecht. The above results are our main observations. In a similar way we calculated *resUAr* and *resUArCC* in relation to the UAr-regression line. In order not to complicate the discussion we refer to Fig 26 for a further explanation of the different residuals and leave the conclusions to the reader.

We conclude this section with a remark about population densities. Are multi-governance urban agglomeration less densely populated than one-governance urban agglomeration? If so, it could be an explanation why most of the urban agglomerations in the Netherlands are characterized by underperformance. But this is not the case. A detailed analysis in our earlier study [14] showed that population density does not relate to the measured residuals. In other words, over- and underperformance of cities as compared to the expected GUP values cannot be attributed to population density. The Rotterdam agglomeration has 8 municipalities, and the overall density is 2680 inhabitants/km$^2$. Compare this with two German harbor cities, Hamburg and Bremen, both are kreisfrei, i.e., their agglomerations consist of only one municipality. The densities for these *one-governance* urban regions are for Hamburg 2304 inhabitants/km$^2$ and for Bremen 1682 inhabitants/km$^2$, well below the *multi-governance* Rotterdam agglomeration. The densities of the multi-governance The Hague agglomeration (4 municipalities) and the multi-governance Leiden agglomeration (5 municipalities) are 3656 and 2775 inhabitants/km$^2$, considerably higher than the population densities of Hamburg and Bremen or of Cologne and Düsseldorf with 2542 and 2744 inhabitants/km$^2$, respectively. Again, we conclude that population densities are most probably not a main factor in the difference of GUP between one-governance and multi-governance urban areas.

|  | resAll | resCC | resUAg | resUAgCC | resUAr | resUArCC |
|---|---|---|---|---|---|---|
| **Rotterdam** | -0.17 | *-0.11* |  |  |  |  |
| Schiedam | 0.06 | *-0.05* |  |  |  |  |
| Nissewaard | -0.52 | *-0.62* |  |  |  |  |
| Vlaardingen | -0.38 | *-0.50* |  |  |  |  |
| Capelle aan den IJssel | 0.41 | *0.28* |  |  |  |  |
| Barendrecht | 0.39 | *0.24* |  |  |  |  |
| Maassluis | -0.39 | *-0.57* |  |  |  |  |
| Krimpen aan den IJssel | 0.01 | *-0.18* | *-0.17* | *-0.38* |  |  |
| Ridderkerk | 0.28 | *0.13* |  |  |  |  |
| Hellevoetsluis | -0.43 | *-0.60* |  |  |  |  |
| Albrandswaard | 0.31 | *0.11* |  |  |  |  |
| Brielle | 0.44 | *0.20* |  |  |  |  |
| Westvoorne | 0.00 | *-0.25* |  |  |  |  |
| Lansingerland | 0.07 | *-0.06* |  |  | -0.11 | -0.46 |
|  |  |  |  |  |  |  |
| **Den Haag** | -0.29 | *-0.25* |  |  |  |  |
| Leidschendam-Voorburg | -0.54 | *-0.65* |  |  |  |  |
| Rijswijk (ZH.) | 0.66 | *0.51* |  |  |  |  |
| Wassenaar | 0.53 | *0.32* | *-0.15* | *-0.34* |  |  |
| Zoetermeer | -0.07 | *-0.14* |  |  |  |  |
| Westland | 0.25 | *0.16* |  |  |  |  |
| Delft | 0.06 | *-0.03* |  |  |  |  |
| Pijnacker-Nootdorp | -0.51 | *-0.65* |  |  |  |  |
| Midden-Delfland | 0.19 | *-0.04* |  |  | -0.12 | -0.47 |
|  |  |  |  |  |  |  |
| **Leiden** | 0.09 | *0.01* |  |  |  |  |
| Leiderdorp | 0.02 | *-0.18* |  |  |  |  |
| Voorschoten | -0.57 | *-0.77* |  |  |  |  |
| Oegstgeest | -0.10 | *-0.31* |  |  |  |  |
| Zoeterwoude | 1.51 | *1.22* | *-0.11* | *-0.25* |  |  |
| Teylingen | 0.14 | *-0.04* |  |  |  |  |
| Noordwijk | 0.38 | *0.18* |  |  |  |  |
| Noordwijkerhout | 0.20 | *-0.04* |  |  |  |  |
| Katwijk | -0.35 | *-0.48* |  |  | -0.14 | -0.45 |
|  |  |  |  |  |  |  |
| **Dordrecht** | -0.08 | *-0.16* |  |  |  |  |
| Zwijndrecht | 0.14 | *-0.02* |  |  |  |  |
| Papendrecht | 0.32 | *0.14* |  |  |  |  |
| Sliedrecht | 0.63 | *0.42* | *-0.10* | *-0.25* |  |  |
| Hardinxveld-Giessendam | 0.38 | *0.15* |  |  |  |  |
| Alblasserdam | 0.31 | *0.09* |  |  |  |  |
| Hendrik-Ido-Ambacht | -0.24 | *-0.43* |  |  | -0.06 | -0.36 |

*Table 6  Set of residuals for the PZH urban agglomerations and urban areas.*

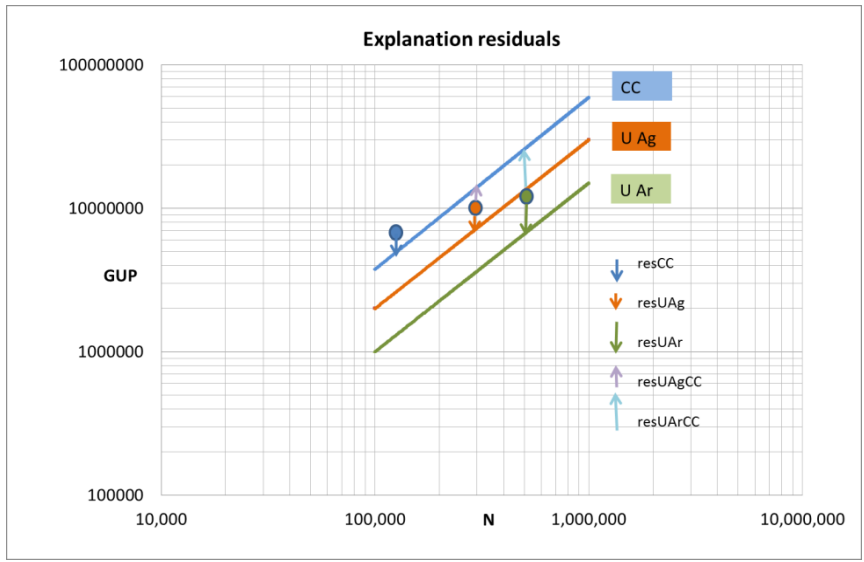

*Fig 26 Explanation of the residuals in relation to the regression lines of the 21 major cities, their agglomerations, and their urban areas.*

*4.3.3. Some first observations in relation to municipal reform*

Finally, we focused our analysis on the 10 PZH municipalities that have been reformed in the period 2000-2009 and for which the municipal reform involved an increase of at least 10% in population as compared to the largest city/town in the new municipality. We take 2010 as the last year of our analysis in order to give the newly formed municipalities time to adjust to the new situation. We calculated the ratio of the GUP for the year 2016 and the GUP for the year 2013. We find for the 10 reformed municipalities 1.14 [sd=0.07], and for the 50 not-reformed municipalities 1.08 [sd=0.05], a difference of about 6% to the advantage of the reformed municipalities.

## 5. Conclusions and policy implications

In most earlier work on urban scaling the 'cities' are in fact larger agglomerations around central cities. It is emphasized [3] that these agglomerations are socioeconomic units and therefore the defining feature of cities, this in contrast to administrative definitions which are regarded as more arbitrary. We however argue that governance structure of urban agglomerations does have a great influence on the socio-economic strength of cities and their agglomerations. This governance structure often has very longstanding and deep historical, political and social grounds that are frequently the basis of emotional attitudes against the central city. Emotions are related to issues such as identity, supposed threats (lower income housing, higher taxes, the loss of green space and local facilities), and proximity of local authorities. Particularly identity is a very strong emotion. Recently Fukuyama stated that for the recognition of our identity we are prepared to sacrifice safety, justice and democracy [32]. Given the findings of this study, we could add: sacrifice a piece of prosperity.

In this study we have investigated the scaling behavior of all Danish municipalities, all German kreisfreie cities (major cities of which the surrounding urban region belongs to the municipality of the city) and all Kreise (regions around smaller cities consisting of several municipalities). For the Netherlands we analyzed the group of major cities including their urban agglomerations and urban areas; and all municipalities in the Province of South-Holland (PZH).

In the case of Denmark we analyzed the scaling of larger cities, municipalities within the Copenhagen agglomeration, and municipalities in rural areas. We also distinguished between municipalities with high and low centrality. In all cases superlinear urban scaling of the gross urban/municipal product with population was found with scaling exponents between 1.14 and 1.24. Also in the case of Germany we find significant superlinear scaling of the gross urban product with population size with exponents up to 1.33. Moreover, our analysis shows that urban regions with *one municipality* (one-governance, kreisfreie cities) perform better than urban regions with *fragmented governance* structures (more than one municipality, multi-governance). We find a strong relation between the measured residuals of the scaling equations and the socio-economic position of a cities as measured with a combination of different socioeconomic indicators.

In both Dutch cases -the 21 major cities with their agglomerations and all municipalities in the Province of Zuid-Holland (PZH)- again significant superlinear scaling is measured with exponents up to 1.28. Our earlier observation that one-governance urban areas perform better than multi-governance urban areas is confirmed and this is in line with the findings for Germany in this study.

Undoubtedly, the independent municipalities within urban agglomerations will have socioeconomic connections. But this does not mean that this multi-governance structure within these agglomerations has a strong cohesiveness and synergy resulting in an optimal social, economic and cultural coherence. Quite the contrary, the independent, autonomous municipalities within urban agglomerations and urban areas each have their own political and social agenda. Even a medium-sized compact urban area may consist of ten autonomous municipalities with in total about 400,000 inhabitants. Every four years there are new municipal elections which may involve a complete change of political orientation. This often results in new policy making in which previous collaboration agreements and partnerships within the agglomeration may be revised or even eliminated thereby eroding the culture of mutual confidence. As a consequence, urban agglomerations may suffer considerably for many decades from the lack of vigor and perseverance in the realization of infrastructural, social, cultural and economic (particularly industrial business areas) facilities.

Our observations in this study lead to challenging conclusions about the importance of a one-municipality instead of a multi-municipality governance in major urban regions. A coherent governance of major cities and their agglomerations may create more effective social interactions which reinforce economic and cultural activities generating a substantial wealth benefit. Even in the case that not most, or all, of the differences in performance between central cities and their urban agglomerations and urban areas can be explained by incoherent governance, then still a substantial part of the expected benefits would generate a significant increase of wealth and disposable resources. For instance, if the benefit would be only 10% of the expected value, then we are still talking about a hundred million Euros per medium-sized city, which means thousands of jobs.

Inter-municipal collaboration is meant to improve the relations between central cities and their suburban municipalities, but it is not known how far the improvements would go if the urban area would change into a one-governance structure. This study provides strong indications for the benefits of a one-governance structure. These indications are in accordance with the findings of the OECD study on the role of urban governance in making cities more productive [13]. Increasing size of cities lead to greater professionalization, higher organizational specialization and increased administrative capacity. But above all, increasing means a disproportional increase of socioeconomic strength. Just like the recent Dutch study on urban regions as drivers of economic growth [34] our work underlines the importance of more effective governance in urban areas. Given that major cities are important drivers of a country's socio-economic development, *the lack of coherent urban governance may severely hamper developments in national wealth*. US researchers concluded in a recent paper: " ……The Netherlands could become even richer simply by growing their cities further" [12, p. 10].

## Acknowledgements

This research is part of a project on the relation between spatial scale and governance synergy. The author thanks his colleagues in the project team, Willem Goedhart, Gerwin van der Meulen (both Decisio Economic Consulting, Amsterdam), Pieter Tordoir (University of Amsterdam) and Frank van Oort (Erasmus University, Rotterdam) for inspiring discussions and valuable comments. This project is financially supported by the Province of South-Holland (PZH).

*Disclaimer*

*The content of this publication does not reflect the official opinion of the Province of South-Holland. Responsibility for the data collection and data analysis as well as for the conclusions based on the empirical results lies entirely with the author.*

## References

1. Florida R (2004). *Cities and the creative class*. New York: Routledge.

2. Bettencourt LMA, Lobo J, Helbing D, Kühnert C, West GB (2007). Growth, innovation, scaling, and the pace of life in cities. *Proc Natl Acad Sci USA* 104, 17: 7301-7306.

3. Bettencourt LMA, Lobo J, Strumsky D, West GB (2010). Urban Scaling and Its Deviations: Revealing the Structure of Wealth, Innovation and Crime across Cities. *PLoS ONE* 5, 11, e13541.

4. Lobo J, Bettencourt LMA, Strumsky D, West GB (2013). Urban scaling and the production function for cities. *PLoS ONE* 8 (3): e58407.

5. Schläpfer M, Bettencourt LMA, Grauwin S, Raschke M, Claxton R, Smoreda Z, et al.(2014). The scaling of human interactions with city size. *Journal of the Royal Society Interface* 11, 98: 20130789.

6. Arbesman S, Kleinberg JM, Strogatz SH (2009). Superlinear scaling for innovation in cities. *Phys. Rev* E 68: 066102.

7. Bettencourt LMA, Lobo J, Strumsky D (2007). Invention in the city: Increasing returns to patenting as a scaling function of metropolitan size. *Research Policy* 36: 1007-120.

8. Nomaler Ö, Frenken K, Heimeriks G (2014) On Scaling of Scientific Knowledge Production in U.S. Metropolitan Areas. *PLoS ONE* 9(10): e110805.

9. Van Raan AFJ (2013). Universities Scale Like Cities. *PLoS ONE* 8(3): e59384.

10. Holland JH (1995). *Hidden Orders. How Adaptation Builds Complexity*. New York: Basic Books.

11. Bettencourt LMA (2013). The Origins of Scaling in Cities. *Science 340*: 1438-1441.

12. Bettencourt LMA, Lobo J (2016). Urban scaling in Europe. *J. of the Royal Society Interface* 13: 2016.0005.

13. OECD (2012). Redefining 'urban': a new way to measure metropolitan areas. Paris: OECD Publishing. For data see:
http://ec.europa.eu/regional_policy/sources/docgener/focus/2012_01_city.pdf and
http://www.oecd.org/cfe/regional-policy/functionalurbanareasbycountry.htm.

14. van Raan AFJ, van der Meulen G, Goedhart W (2016). Urban Scaling of Cities in the Netherlands. *PLoS ONE* 11(1): e0146775.

15. Ahrend, R., Farchy E., Kaplanis, I., Lembcke, A. (2014). *What Makes Cities More Productive? Evidence on the Role of Urban Governance from Five OECD Countries*. OECD Regional Development Working Papers, No. 2014/05. Paris: OECD Publishing.

16. Pumain D (2004). *Scaling laws and urban systems*. SFI Working Paper 2004-02-002, Santa Fe institute. Available online at: http://www.santafe.edu/media/workingpapers/04-02-002.pdf.

17. Pumain D (2012). Urban Systems Dynamics, Urban Growth and Scaling Laws: The Question of Ergodicity. In Portugali J, Meyer H, Stolk E, Tan E (eds.): *Complexity theories of cities have come of age: an overview with implications to urban planning and design.* Heidelberg, Berlin: Springer.

18. Blom-Hansen J (2010). Municipal Amalgamations and Common Pool Problems: The Danish Local Government Reform in 2007. *Scandinavian Political Studies* 33(1): 51-73.

19. Hansen SW (2007). Towards Genesis or the Grave: *Financial Opportunism in the Face of Local Government Mergers in Denmark*. Conference paper, 17th Nordiske Kommunalforskerkonference, Göteborg, Sweden.

20. Hinnerich BT (2009) Do merging local governments free ride on their counterparts when facing boundary reform? *Journal of Public Economics* 93: 721-728.

21. Jordahl H, Liang C (2010). Merged municipalities, higher debt: on free-riding and the common pool problem in politics. *Public Choice* 143: 157-172.

22. Hansen SW (2014). Common pool size and project size: an empirical test on expenditures using Danish municipal mergers. *Public Choice* 159: 3–2.

23. Saarimaa T, Tukiainen J (2015). Common pool problems in voluntary municipal mergers. *European Journal of Political Economy* 38: 140–152.

24. Hansen SW, Houlberg K, Pedersen LH (2014). Do Municipal Mergers Improve Fiscal Outcomes? *Scandinavian Political Studies* 37(2): 196-214.

25. Lassen DD, Serritzlew S (2011). Jurisdiction Size and Local Democracy: Evidence on Internal Political Efficacy from Large-scale Municipal Reform. *American Political Science Review* 105: 238-258.

26. Rouse P, Putterill M (2005). Local government amalgamation policy: A highway maintenance evaluation. *Management Accounting Research* 16: 438-463.

27. Lüchinger S, Stutzer A (2002). Skalenerträge in der öffentlichen Kernverwaltung: eine empirische Analyse anhand von Gemeindefusionen. *Swiss Political Science Review* 8: 27-50.

28. Moisio A, Uusitalo R (2013). The impact of municipal mergers on local public expenditures in Finland. *Public Finance and Management* 13: 148-166.

29. Reingewertz Y (2012). Do Municipal Amalgamations Work? Evidence from Municipalities in Israel. *Journal of Urban Economics* 72: 240–251.

30. Prognos Zukunft Atlas (2016). Retrieved from: https://www.prognos.com/publikationen/zukunftsatlas-r-regionen/zukunftsatlas-r-2016/.

31. Arshad S, Hu S, Ashraf BN (2018). Zipf's law and city size distribution: A survey of the literature and future research agenda. *Physica A* 492 75–92.

32. Market G, van Woerkens C (2017). *Atlas voor Gemeenten 2017*. Nijmegen: VOC Uitgevers. See also https://www.atlasvoorgemeenten.nl/de-atlas/de-atlas.

33. Fukuyama F (2018). Identity: The Demand for Dignity and the Politics of Resentment. New York: Farrar, Straus and Giroux.

34. Raspe O, van den Berge M, de Graaff T (2017). *Stedelijke regio's als motoren van economische groei. Wat kan beleid doen?* Den Haag: Planbureau voor de Leefomgeving (PBL), PBL publication nr 2901.

Appendix:

Calculation of the residuals

We calculated the residuals of the power-law scaling of the gross urban product with population for the analysis of the real performance as compared to the expected value. The mathematical procedure is as follows.

A power-law relation between for instance the gross urban product ($G$) and population ($P$) can be written as:

$$G(P) = aP^{\beta} \tag{A1}$$

We find empirically (as an example see Fig 1, upper part) the value 66.19 for the coefficient *a* and 1.14 for the power-law exponent ß.

Denoting the observed value of the gross urban product for each specific city with $G_i$, we calculate the residuals $\xi_i$ of the scaling distribution of each city as follows [3, 14]:

$$\xi_i = ln[G_i/G(P)] = ln[G_i/aP^{\beta}] \tag{A2}$$

The residuals are also used to test the heteroscedasticity of the data, we refer to our earlier paper [14].